\documentclass[aps,prl,reprint]{revtex4-1}
\usepackage{amsmath}
\usepackage{graphicx}
\usepackage{units}  
\usepackage{xspace}
\usepackage{subfigure}
\usepackage{hyperref}

\newcommand{\ve}{\varepsilon}
\newcommand{\mb}{\mathbf}
\newcommand{\tb}{\textbf}
\newcommand{\beq}{\begin{equation}}
\newcommand{\eeq}{\end{equation}}
\newcommand{\bea}{\begin{eqnarray}}
\newcommand{\eea}{\end{eqnarray}}

\usepackage[usenames]{color}

\begin{document}

\bibliographystyle{apsrev}
 
\title{Unusual Frequency of Quantum Oscillations in Strongly Particle-Hole Asymmetric Insulators}

\date{\today}
\author{Hridis K. Pal}
   \email{hridis.pal@physics.gatech.edu}
\affiliation{LPS, CNRS UMR 8502, Universit\'{e} Paris-Sud, Universit\'{e} Paris-Saclay, 91405  Orsay Cedex, France}

\begin{abstract} 
Quantum oscillations, conventionally thought to be a metallic property, have recently been shown to arise in certain kinds of insulators, with properties very different from those in metals.
All departures from the canonical behavior found so far arise only in the amplitude and the phase but not in the frequency. Here I show that such robustness in the behavior of the frequency is only valid for a particle-hole symmetric insulator; in a strongly particle-hole asymmetric insulator, de Haas-van Alphen oscillations (oscillations in magnetization and susceptibility) and Shubnikov-de Haas oscillations (oscillations in the density of states) exhibit different frequencies, with the frequency of the latter changing with temperature. I demonstrate these effects with numerical calculations on a lattice model, and provide a theory to account for the unusual behavior.
\end{abstract}

\maketitle 

A direct manifestation of Landau quantization in a magnetic field in metals is the appearance of quantum oscillations. 
Oscillations arise due to Landau levels crossing the Fermi level periodically as the field $B$ is changed, 
and are periodic in $1/B$. The leading harmonic of the oscillating part of some physical observable $Q$ can be generically written as
\beq
Q_{osc}=A\mathrm{cos}\left[\frac{F}{B}+\phi\right],
\label{eqmetal}
\eeq
where $A$ is the amplitude, $F$ is the frequency, and $\phi$ is the phase. According to the canonical Lifshitz-Kosevich theory \cite{sho}, temperature $T$ modifies only the amplitude, in a universal way that applies to all physical quantities, i.e., it is quantity independent. On the other hand the phase $\phi$ depends on the quantity being measured but not on $T$. In contrast to $A$ and $\phi$, the frequency $F$ is independent of both $T$ and the quantity being studied. Thus, $A\rightarrow A(T,B)$, $\phi\rightarrow\phi_Q$, and $F\rightarrow F$.

\begin{figure}
\includegraphics[angle=0,width=0.99\columnwidth,trim={0 4cm 0 0},clip]{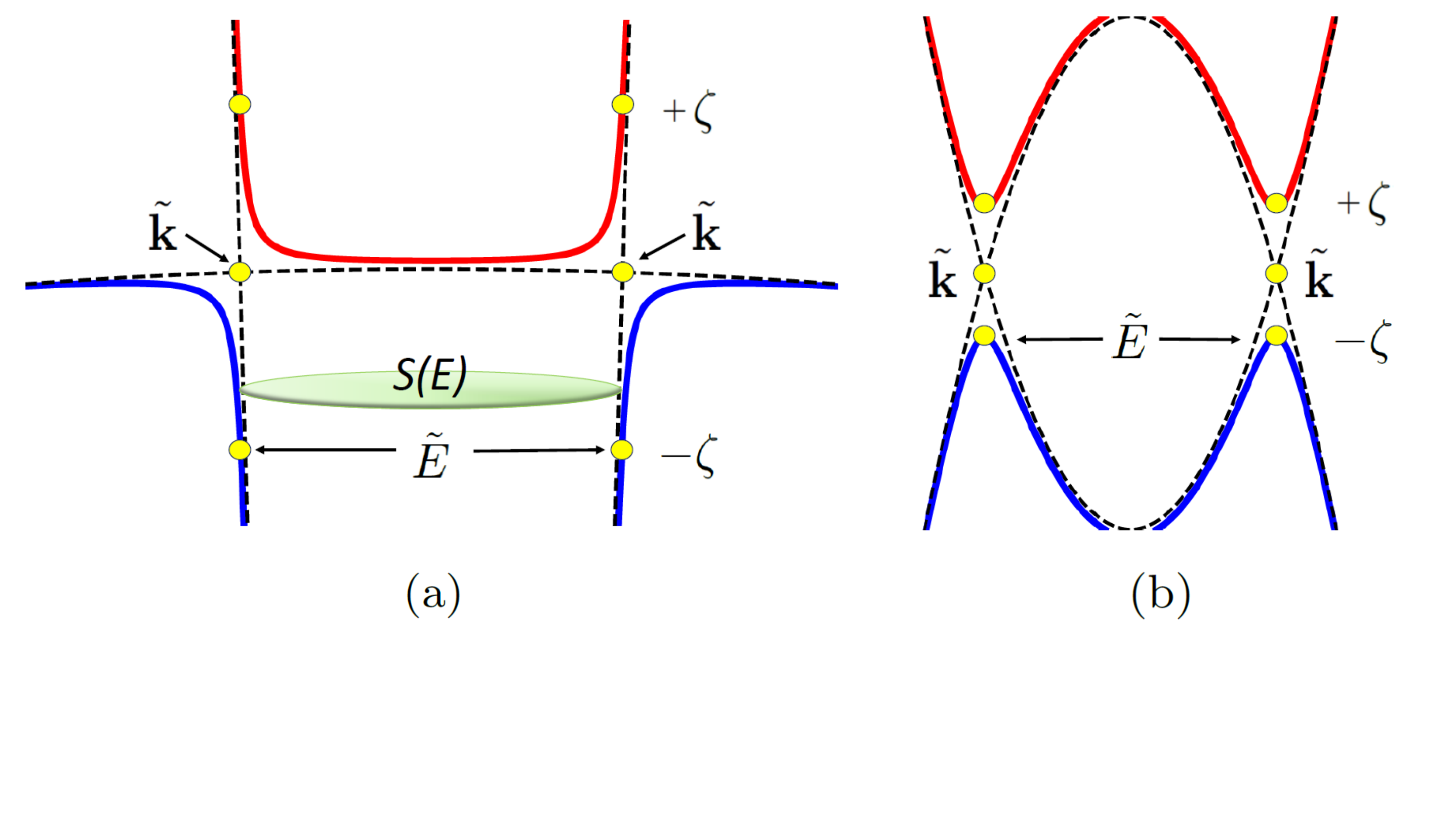}
\caption{Schematic band diagram from Eq.~(\ref{ham}) after hybridization: (a) strongly particle-hole asymmetric and (b) particle-hole symmetric. Unhybridized bands  are shown in dashes, intersecting at $\tilde{\mb{k}}$. At $T=0$, when $\mu=0$ is in the gap, oscillations in both cases originate from $\tilde{E}=E(\tilde{\mb{k}})=-\zeta$. However, while in (a) the region in $[-\zeta,\zeta]$ is partially gapped with the gap $\Delta_g\ll 2\zeta$, in (b) it is completely gapped with the gap $\Delta_g=2\zeta$. Within $[-\zeta,\zeta]$, when $\mu$ is in the band region in (a), there are unusually two sources of oscillations, one from $\mu$ and one from $\tilde{E}$, see Eq.~(\ref{omegadec}) for explanation. This does not arise in (b). Additionally, the band mass and the area of the orbit $S(E)$ change rapidly in this interval in (a) unlike in (b).}
\label{fig1}
\end{figure}

Recently, inspired by experimental observations \cite{tan} of quantum oscillations in SmB$_6$, a Kondo insulator, several theoretical studies have considered the possibility of oscillations without a Fermi surface \cite{kno,zha,pal1,pal2,kno2,kum,fri,bal,rus}. It has been shown that oscillations can arise in certain kinds of insulators, with properties very different from those in metals.
Temperature not only modifies the amplitude but also the phase and has a different dependence compared to metals.
Further, the dependence is no longer universal for all quantities: oscillations in thermodynamic quantities such as magnetization and susceptibility---de Haas-van Alphen (dHvA) oscillations---and in those arising from the density of states such as the resistivity---Shubnikov-de Haas (SdH) oscillations---show different behavior. Notably, however, all departures from the canonical behavior arise only in the amplitude and the phase \cite{kno,zha,pal1}; the frequency $F$ continues to behave as in a metal: it is independent of both temperature and the quantity being studied. Thus, $A\rightarrow A_Q(T,B)$, $\phi\rightarrow\phi_Q(T,B)$, and $F\rightarrow F$.

In this Letter, I show that the robustness in the behavior of frequency is valid only in a particle-hole (PH) symmetric insulator; in insulators with strong PH asymmetry, the frequency of oscillations become both quantity and temperature dependent. In particular, dHvA and SdH oscillations show different frequencies, with the frequency of the latter changing with temperature. Thus,  in a strongly PH asymmetric insulator, in addition to $A\rightarrow A_Q(T,B)$, $\phi\rightarrow\phi_Q(T,B)$, one has $F\rightarrow F_Q(T)$. 

\begin{figure}
\centering
\subfigure[]{\includegraphics[width=0.46\columnwidth]{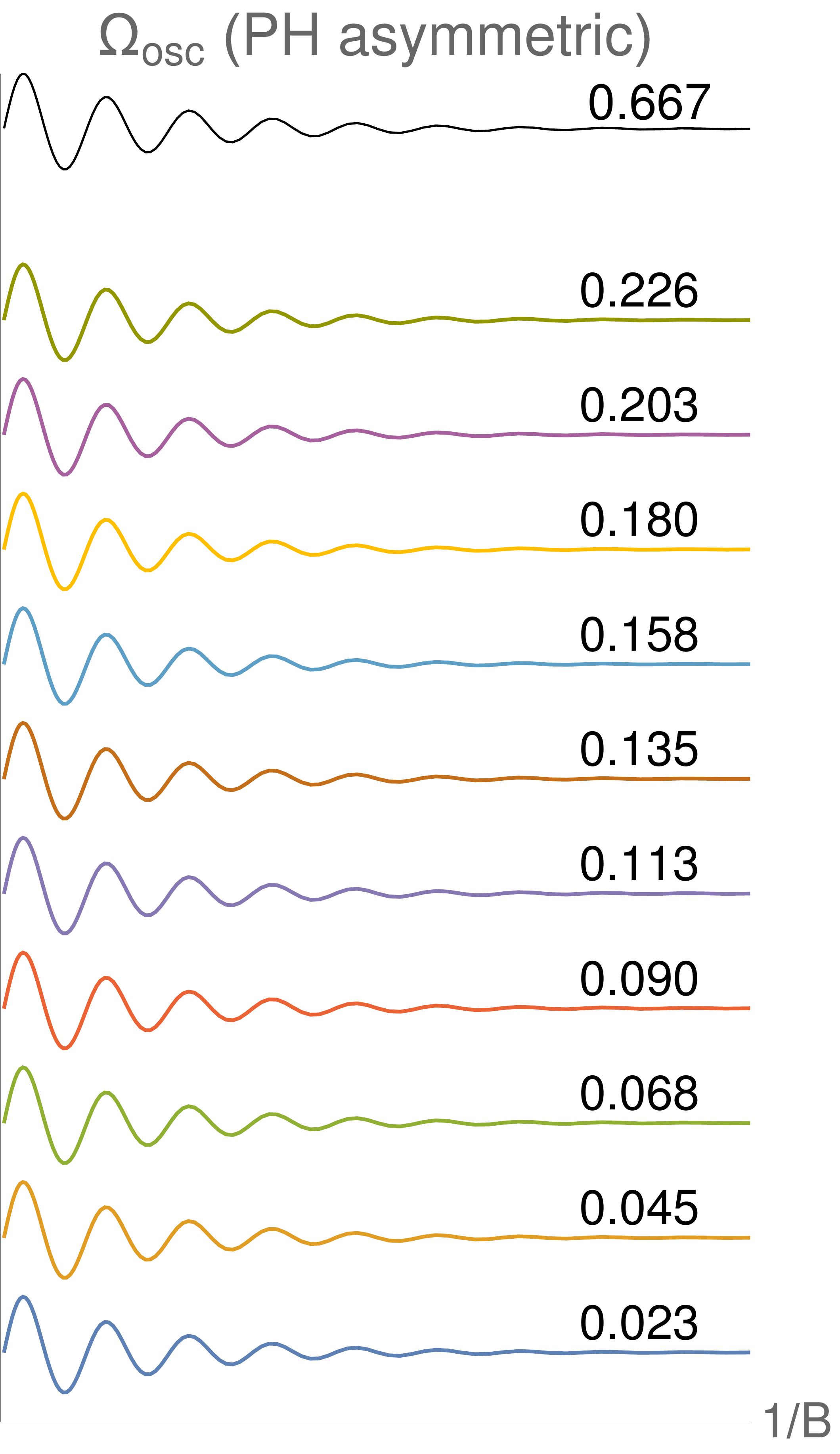}
\label{fig2a}}
\quad
\subfigure[]{\includegraphics[width=0.46\columnwidth]{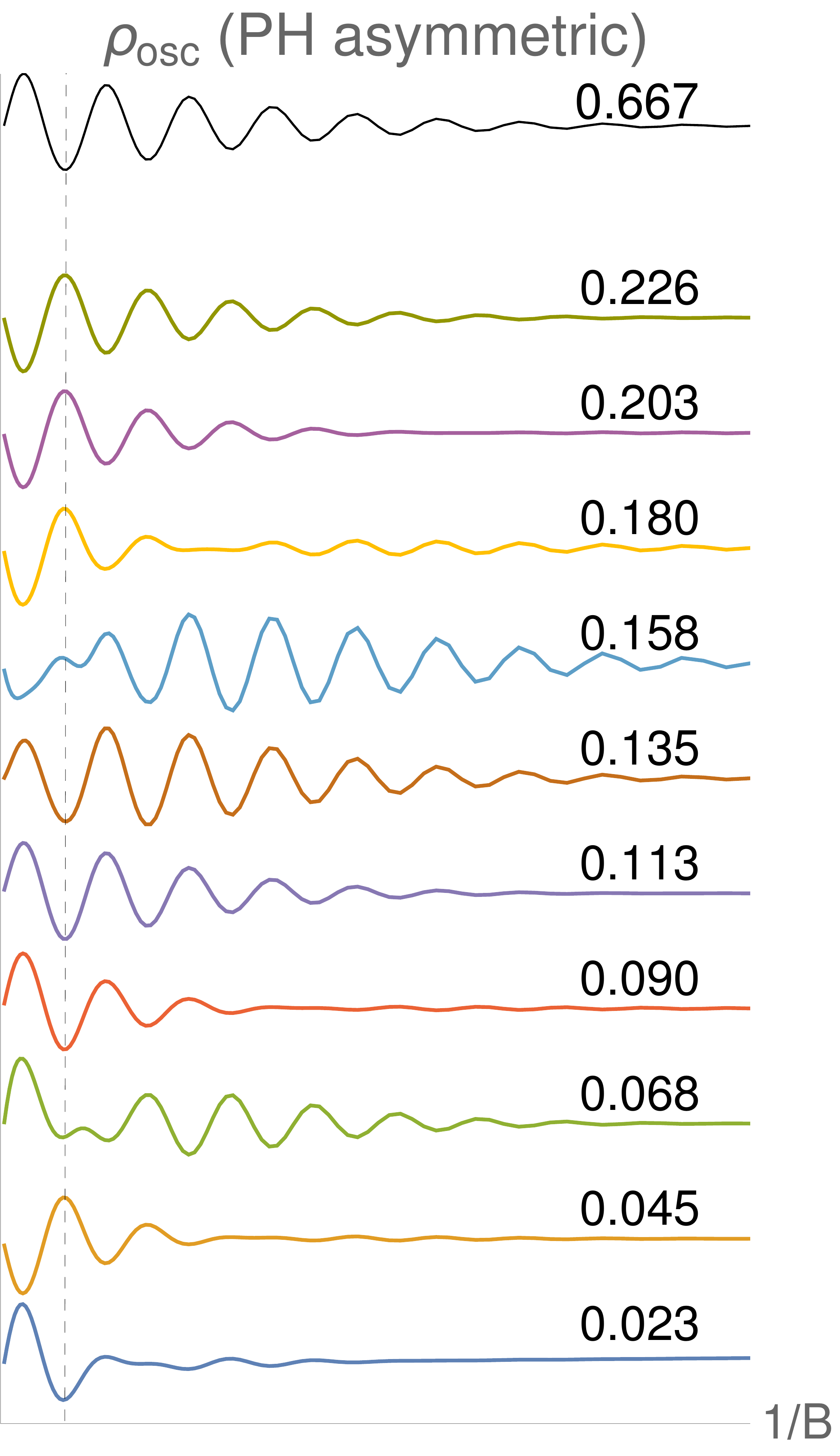}
\label{fig2b}}
\quad
\subfigure[]{\includegraphics[width=0.92\columnwidth]{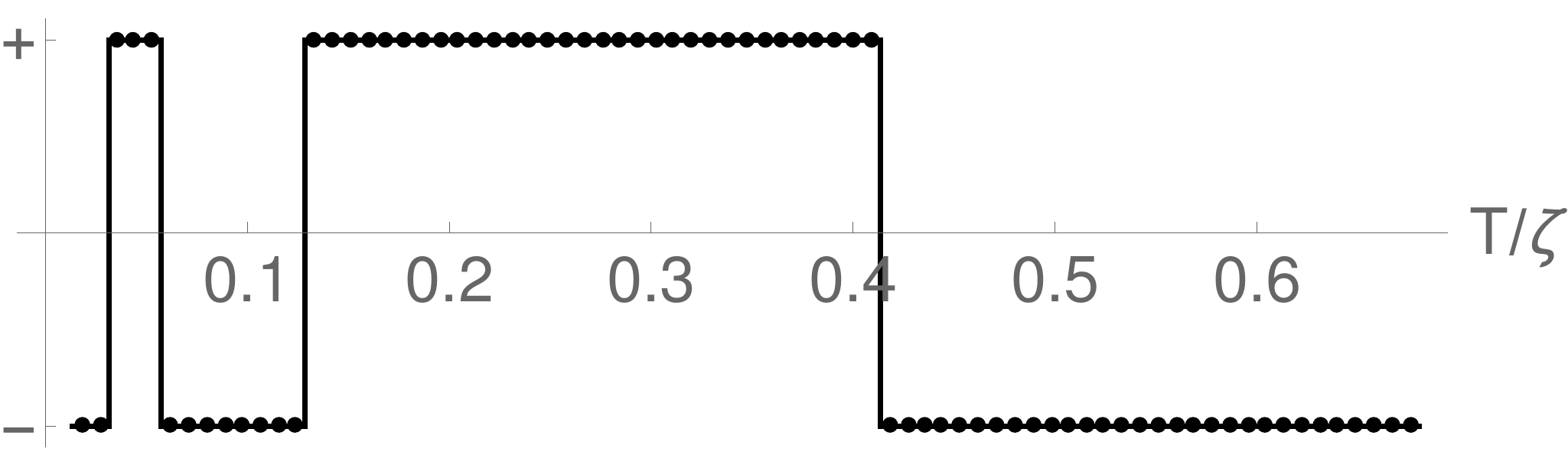}
\label{fig2c}}
\caption{Quantum oscillations numerically calculated for a lattice version of the model in (\ref{ham})---see text for details. Oscillations in (a) grand potential $\Omega$ and (b) density of states $\rho$ for a strongly particle-hole asymmetric case. The curves are rescaled for clarity such that all have the same amplitude within the field interval studied. Corresponding $T/\zeta$ is provided on the right. While (a) shows a single frequency that does not change with $T$, in (b) beats appear signaling more than one frequency. The beat changes with $T$ implying change of frequencies. As a result the phase changes periodically with change in $T$ at a fixed $1/B$. (c) The evolution of phase along the dashed line in (b) with $+$ being a maximum and $-$ being a minimum. The behavior is explained by Eq.~(\ref{res2}).}
\label{fig2}
\end{figure} 

To illustrate these effects, I consider a model with two overlapping bands $\varepsilon_{1,2}$ with masses $m_{1,2}$, different in sign, hybridized by a parameter $\zeta$. The Hamiltonian reads
\beq
H_{\mb{k}}=
\begin{pmatrix}
\ve_1({\mb{k}})-\Delta&\zeta\\
\zeta&\ve_2({\mb{k}})
\end{pmatrix},
\label{ham}
\eeq
where $\Delta$ determines the overlap between the bands before hybridization. 
Let $\mb{k}=\tilde{\mb{k}}$ denote the intersection between the two bands before hybridization. When the bands hybridize, a gap opens up due to avoided crossing. When $|m_2|\gg |m_1|$, the system is strongly PH asymmetric, and when $|m_1|=|m_2|$, it is PH symmetric---see Fig.~\ref{fig1}.

I first numerically demonstrate the effects predicted on a lattice version of the model above: two square lattices intersecting at $1/8$th filling with hopping parameters $t_1$ and $t_2$ are hybridized to open a gap. The chemical potential $\mu$ is placed in the gap such that it would have passed through the band intersection before hybridization. The PH asymmetry is regulated by $m_1/m_2=t_2/t_1$. The energy spectrum in the presence of a magnetic field for such a model can be calculated numerically using the method in \cite{pal1}. Using the spectrum, one can compute the grand potential $\Omega=-T\sum_n\mathrm{ln}(1+e^{(\mu-\ve_n)/T})$.  
Here I present oscillations in $\Omega$ to demonstrate dHvA oscillations (since magnetization $M=-\partial\Omega/\partial B$ behaves similarly as $\Omega$, it suffices to study the latter) and the density of states $\rho=-\partial^2\Omega/\partial\mu^2$ to demonstrate SdH oscillations. In Figs.~\ref{fig2a} and \ref{fig2b}, quantum oscillations in $\Omega$ and $\rho$ are presented for an extremely PH asymmetric case with $t_2/t_1=-0.0001$ and $\zeta/t_1=0.05$. It is seen that at $T<\zeta$, the frequencies of the two quantities do not match: while $\Omega$ shows only one frequency, $\rho$ shows a beat signaling the presence of more than one frequency. Moreover, the beat in $\rho$ changes with $T$, unlike in $\Omega$ where $T$ has no effect on the frequency. At $T\sim\zeta$, the beat in $\rho$ disappears leaving only a single frequency that is same as in $\Omega$. None of these features arise when $t_1=t_2$, confirming that the effects are due to PH asymmetry---see Supplementary Materials \cite{supp}.

\begin{figure}
\includegraphics[angle=0,width=0.99\columnwidth]{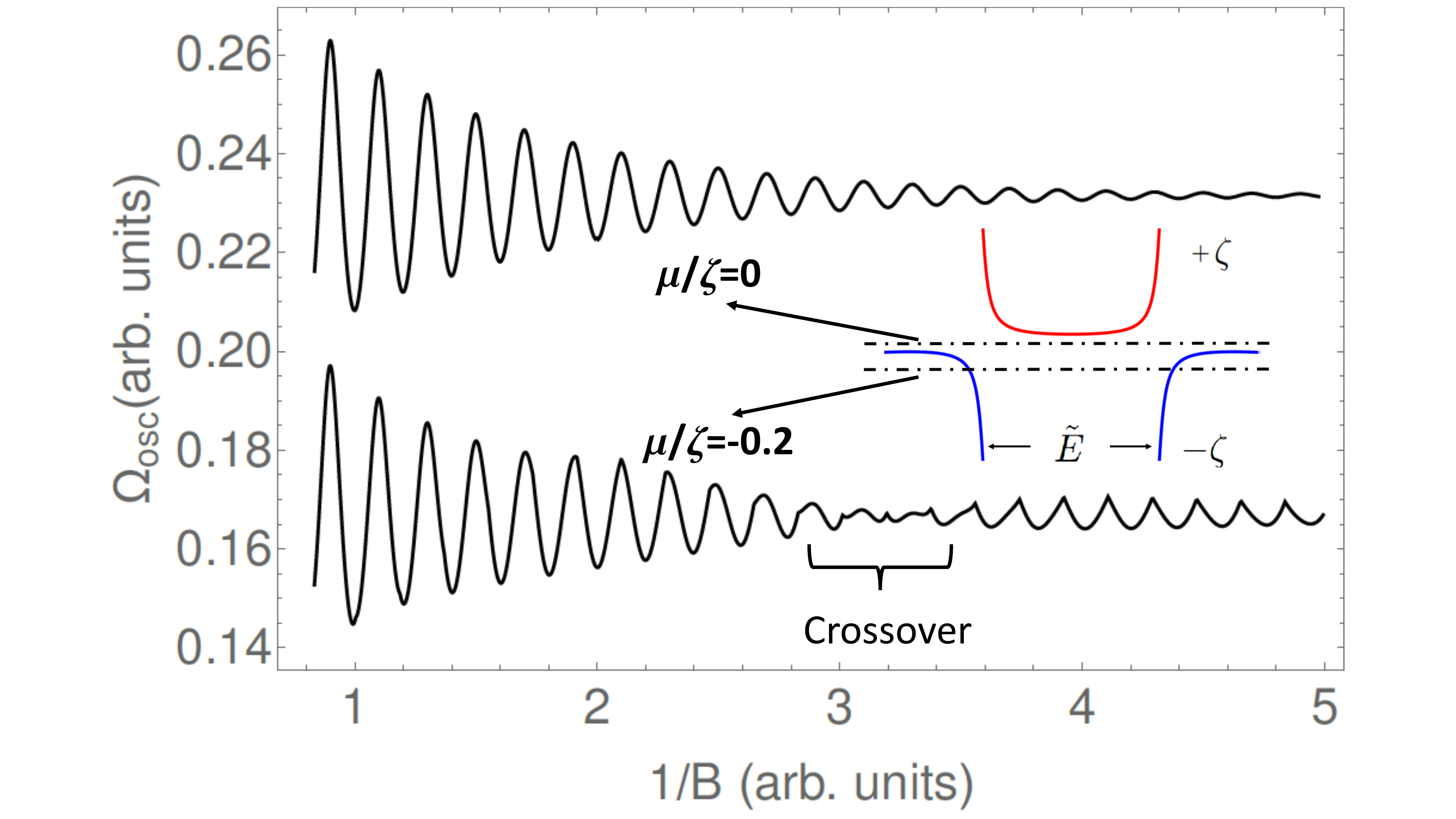}
\caption{Numerically calculated oscillations for the model in (\ref{ham}) with $\varepsilon_{1,2}(\mb{k})=k^2/2m_{1,2}$ with $m_1/m_2=-0.0001$: (a) $\mu$ in the gap and (b) $\mu$ in the valence band but close to the edge. The latter curve follows the  behavior predicted by Eq.~(\ref{omegadec}): At small $1/B$ oscillations are similar to the top curve ($\Omega_{osc}^{un}$) as if $\mu$ is in the gap and oscillations are arising from $\tilde{E}$. As $1/B$ increases, there is a crossover to a different set of oscillations ($\Omega_{osc}^c$) which arises from $\mu$. See Supplementary Materials \cite{supp} for more details.}
\label{fig3}
\end{figure}

The behavior presented above is rather unusual. To facilitate an understanding I use the following formula:
\beq
\Omega_{osc}(\mu,T)=\int_{-\infty}^{\infty}\frac{-\partial f_0(E-\mu,T)}{\partial E}\Omega_{osc}(E,0)dE,
\label{omegatori}
\eeq
where $f_0$ is the Fermi-Dirac function. 
In the regime of interest, $T<\zeta$, where novel features arise, the main contribution to the integral comes from the interval $[-\zeta,\zeta]$ (I put $\mu=0$ for simplicity); therefore, one needs to know $\Omega(E)$ in this interval at $T=0$. As seen in Fig.~\ref{fig1}, this interval in the asymmetric case is qualitatively different from the symmetric case: whereas the symmetric case is totally gapped in this interval, the asymmetric one has both a gapped region and band regions. When $E$ is in the gap, clearly oscillations can not arise from $E$. In Ref.~\cite{pal2} it was shown that oscillations in such cases arise from the sudden change of band slope due to hybridization. This sudden change happens at momentum $\tilde{\mb{k}}$ where the bands were degenerate prior to hybridization; the corresponding energy post  hybridization is $\tilde{E}=E(\tilde{\mb{k}})=-\zeta$ (see Fig.~\ref{fig1}). Thus, oscillations arise from $\tilde{E}=-\zeta$ inside the band. These unconventional (\emph{un}) oscillations, are described by \cite{pal1,pal2}
\beq
\Omega_{osc}^{un}\propto g\left(\frac{\zeta}{\omega_c}\right)\mathrm{cos}[S(\tilde{E})l_B^2+\phi].  \label{omegaunzero}
\eeq
Here, $S(\tilde{E})$ is the area of the $k-$space orbit at $\tilde{E}$ which contributes to the frequency and $l_B=1/\sqrt{eB}$ is the magnetic length ($e$ is the absolute value of the electronic charge and $\hbar$=1) captures the periodicity in inverse field. The form is similar to Eq.~(\ref{eqmetal}) valid for metals, except that oscillations decay with $1/B$ (even at $T=0$) and  $g(\zeta/\omega_c)$ is a function describing it, whose exact form is not required \cite{pal2}. 
Next, if $E$ moves into the band regions [but still within $[-\zeta,\zeta]$---see Fig.~\ref{fig1}(a)], since it is now in the metallic regime, one would expect conventional (\emph{c}) oscillations described by \cite{sho}
\beq
\Omega_{osc}^c(E)\propto\frac{1}{|m(E)|} \mathrm{cos}[S(E)l_B^2+\phi].\label{omegaczero}
\eeq
This is same as Eq.~(\ref{eqmetal}) except that the dependence of the amplitude on the mass $m$, which changes rapidly with $E$, is explicitly stated for future reference. This description, however, turns out to be incomplete. As the field is changed and the Landau levels move, they still encounter the sudden slope change at $\tilde{E}$ before reaching $E$. Thus, on top of the conventional oscillations arising from $E$, unconventional oscillations that were there when $\mu$ was in the gap are also expected. They should be present as long as $E$ is inside $[-\zeta,\zeta]$, and vanish outside this interval. Thus, at $T=0$, inside the band regions in $[-\zeta,\zeta]$ there are, counterintuitively, \emph{two} sources of oscillations: 
\beq
\Omega_{osc}(E)=\Omega_{osc}^c(E)+\Omega_{osc}^{un},
\label{omegadec}
\eeq
Importantly, the two contributions are not of equal strength: for $\zeta/\omega_c\ll 1$ one can expand $g(\zeta/\omega_c)$ to get $|\Omega^{un}_{osc}|\sim|\Omega_{osc}^0|[1-\mathcal{O}(\zeta/\omega_c)]$. On the other hand, $|\Omega^c_{osc}|\sim|\Omega_{osc}^0||m^0/m(E)|$, where the superscript $0$ denotes quantities prior to hybridization ($m^0$ is the lighter mass). Since $m(E)$ is extremely large near the band edge $E=0$ [see Fig.~\ref{fig1}(a)], $|\Omega^c_{osc}|$ is smaller than $|\Omega^{un}_{osc}|$ for small $|E|$. As $E$ approaches $\pm\zeta$, they become comparable. On the other hand, when plotted as a function of $1/B$, two distinct regions of oscillations appear: at smaller $1/B$, $\Omega_{osc}^{un}$ wins and at larger $1/B$, $\Omega_{osc}^c$ wins. The crossover happens at $g(\zeta/\omega_c)\approx m^0/m(E)$. This is verified by numerical calculations shown in Fig.~\ref{fig3}. The two regions have different frequencies: according to Eqs.~(\ref{omegaunzero}) and (\ref{omegaczero}), the ratio of these frequencies $F^c/F^{un}$ is expected to be equal to $S(E)/S(\tilde{E})$. Numerical calculations support this as well---see Supplementary Materials \cite{supp}. 

Although Eq.~(\ref{omegadec}) is an intermediary step in calculating Eq.~(\ref{omegatori}), on its own it is a nontrivial statement with interesting consequences. It predicts that, in spite of being a metal,  when $\mu$ is close to the edge, oscillations with \emph{two} frequencies will appear in \emph{separate} regions of $1/B$. This will show up in dHvA oscillations of magnetization $M=-\partial\Omega/\partial B$. In SdH oscillations of density of states $\rho=-\partial^2\Omega/\partial E^2$, however, only one frequency will show up---the conventional one---since the unconventional part does not depend on $E$ and will drop out on taking the derivative. In experiments these unusual features can be measured by doping a strongly PH-asymmetric insulator slightly so that $\mu$ is pushed into one of the bands.

Returning to Eq.~(\ref{omegatori}), note that the gap itself is quite small, $\sim \zeta^2/\Delta\ll \zeta$. Hence, its contribution to the integral can be neglected (this is valid as long as $T\gg\zeta^2/\Delta$). Then, Eq.~(\ref{omegadec}) describes the entire interval $[-\zeta,\zeta]$ and not just the band regions.  Inserting Eq.~(\ref{omegadec}) in Eq.~(\ref{omegatori}) yields $\Omega_{osc}(T)=\Omega^c_{osc}(T)+\Omega^{un}_{osc}(T)$ and  $\rho_{osc}(T)=-\partial^2\Omega_{osc}/\partial\mu^2=\rho^c_{osc}(T)+\rho^{un}_{osc}(T)$. In the regime of interest, $T<\zeta$, not all terms contribute equally. The integral in Eq.~(\ref{omegatori}) gets its dominant contribution from the vicinity of $E=0$. As argued before, here the conventional contribution is much smaller than the unconventional one. Therefore, to leading order one can approximate $\Omega_{osc}(T)\approx\Omega^{un}_{osc}(T)$. The same, however, does not apply to $\rho_{osc}(T)$. As shown before, here the unconventional part does not contribute at all, leaving only the conventional part: $\rho_{osc}(T)=\rho^c_{osc}(T)$. With these simplifications, I now study each term. 

\begin{figure}
\centering
\subfigure[]{\includegraphics[width=.40\columnwidth]{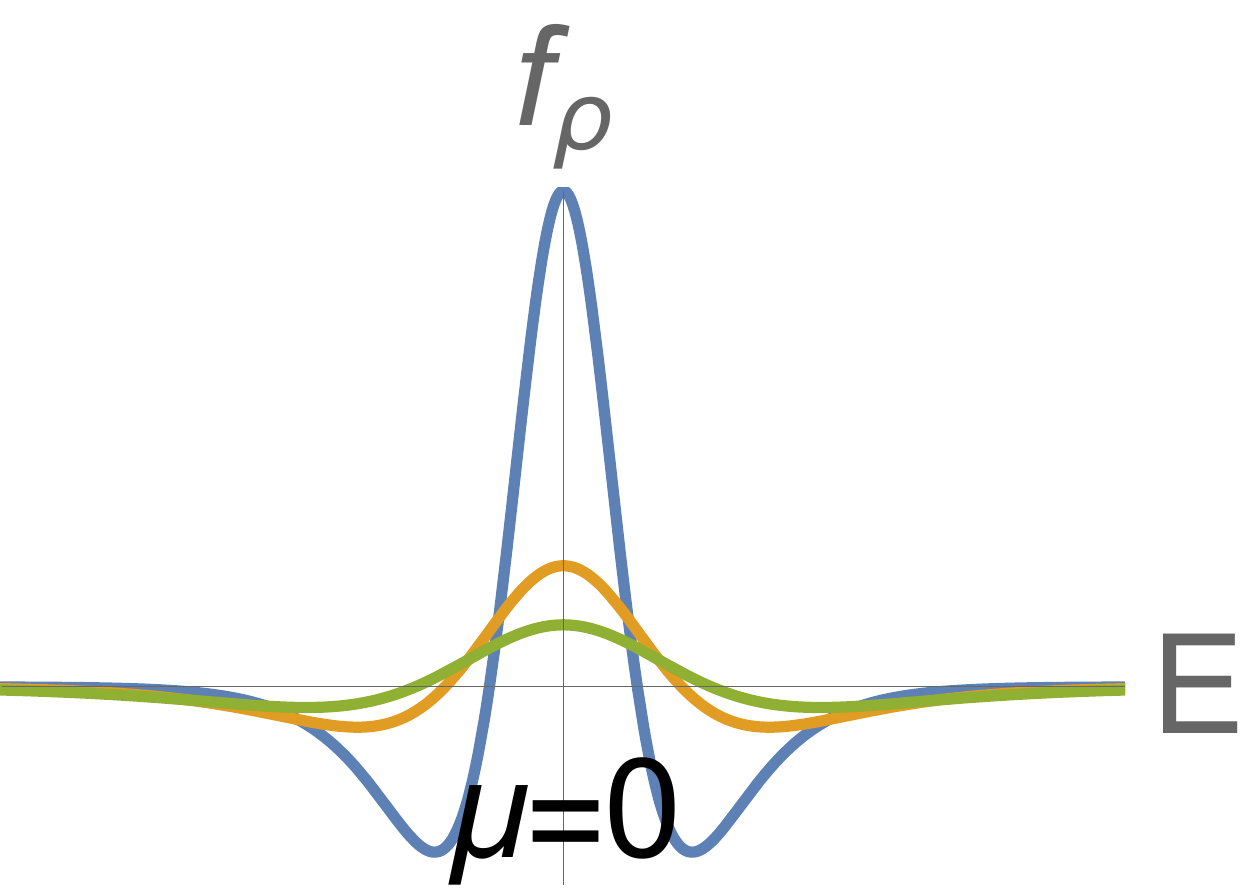}
\label{fig4a}}
\quad
\subfigure[]{\includegraphics[width=.40\columnwidth]{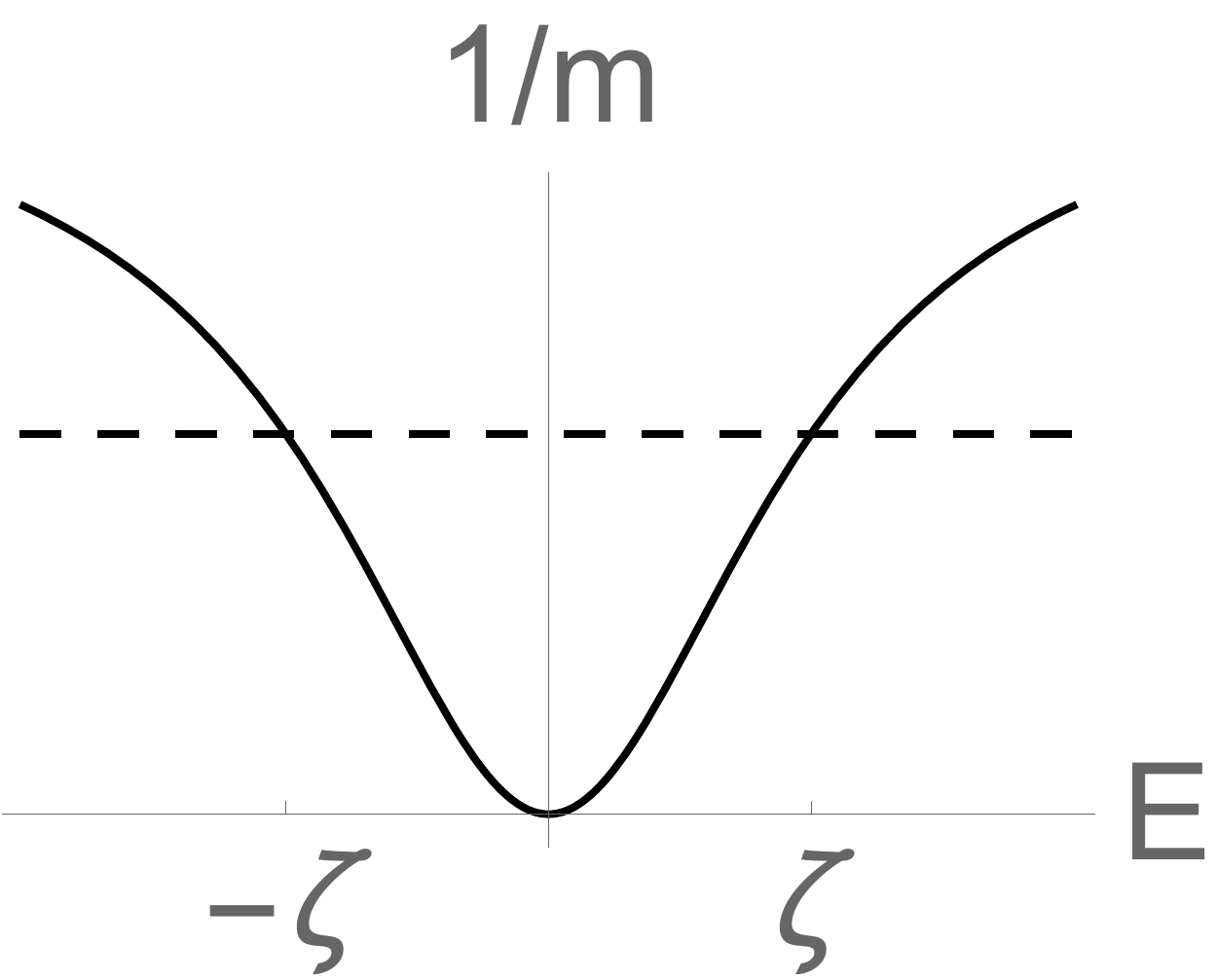}
\label{fig4b}}
\quad
\subfigure[]{\includegraphics[width=.40\columnwidth]{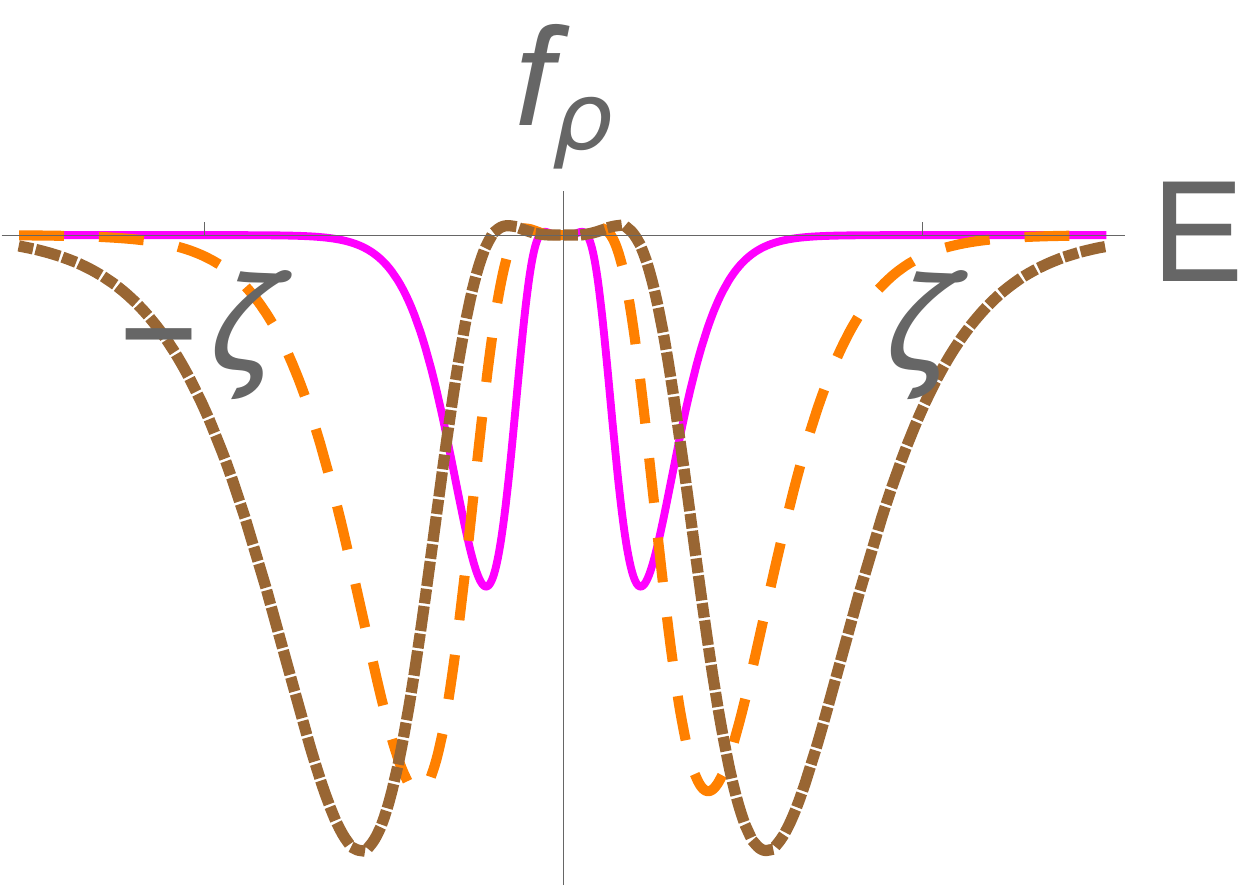}
\label{fig4c}}
\quad
\subfigure[]{\includegraphics[width=.40\columnwidth]{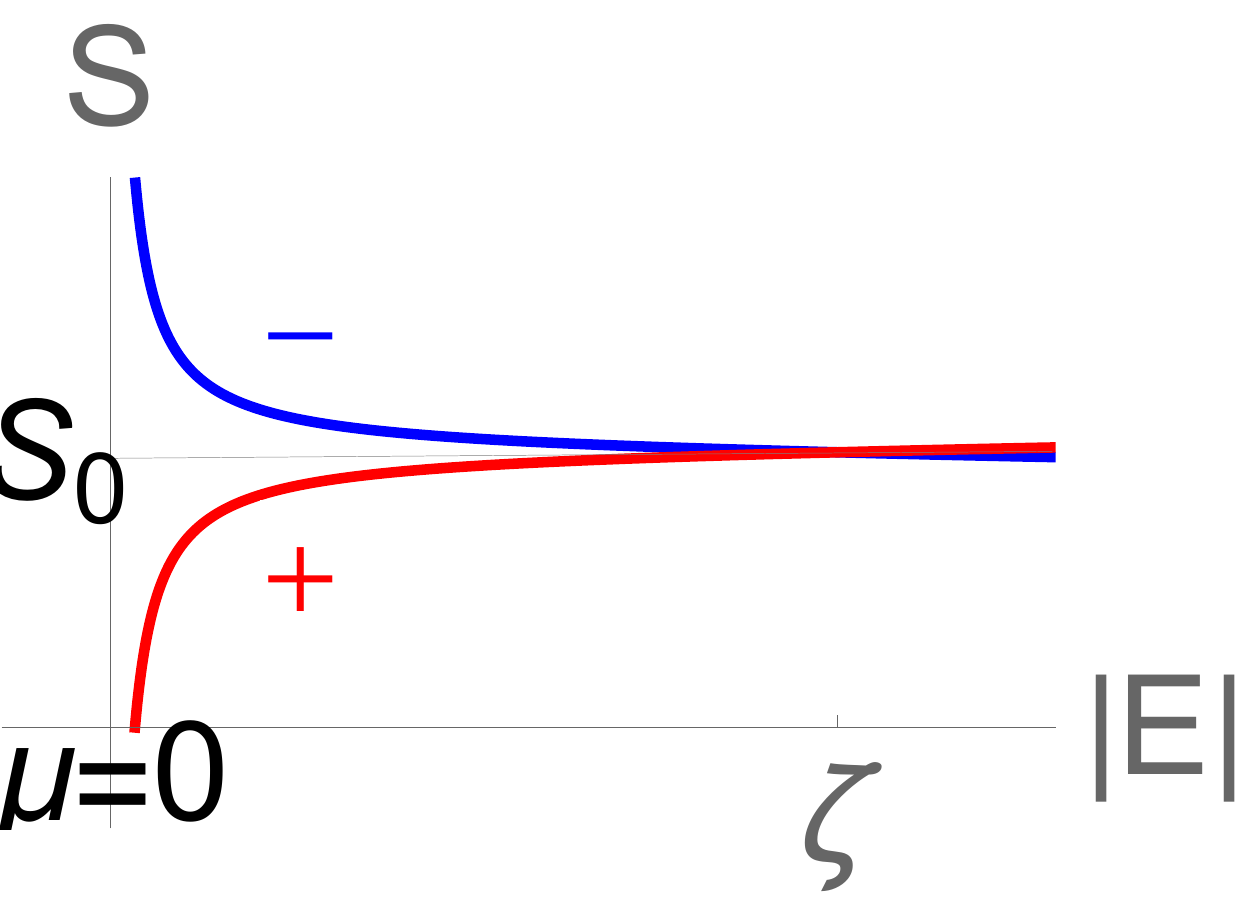}
\label{fig4d}}
\quad
\subfigure[]{\includegraphics[width=.40\columnwidth]{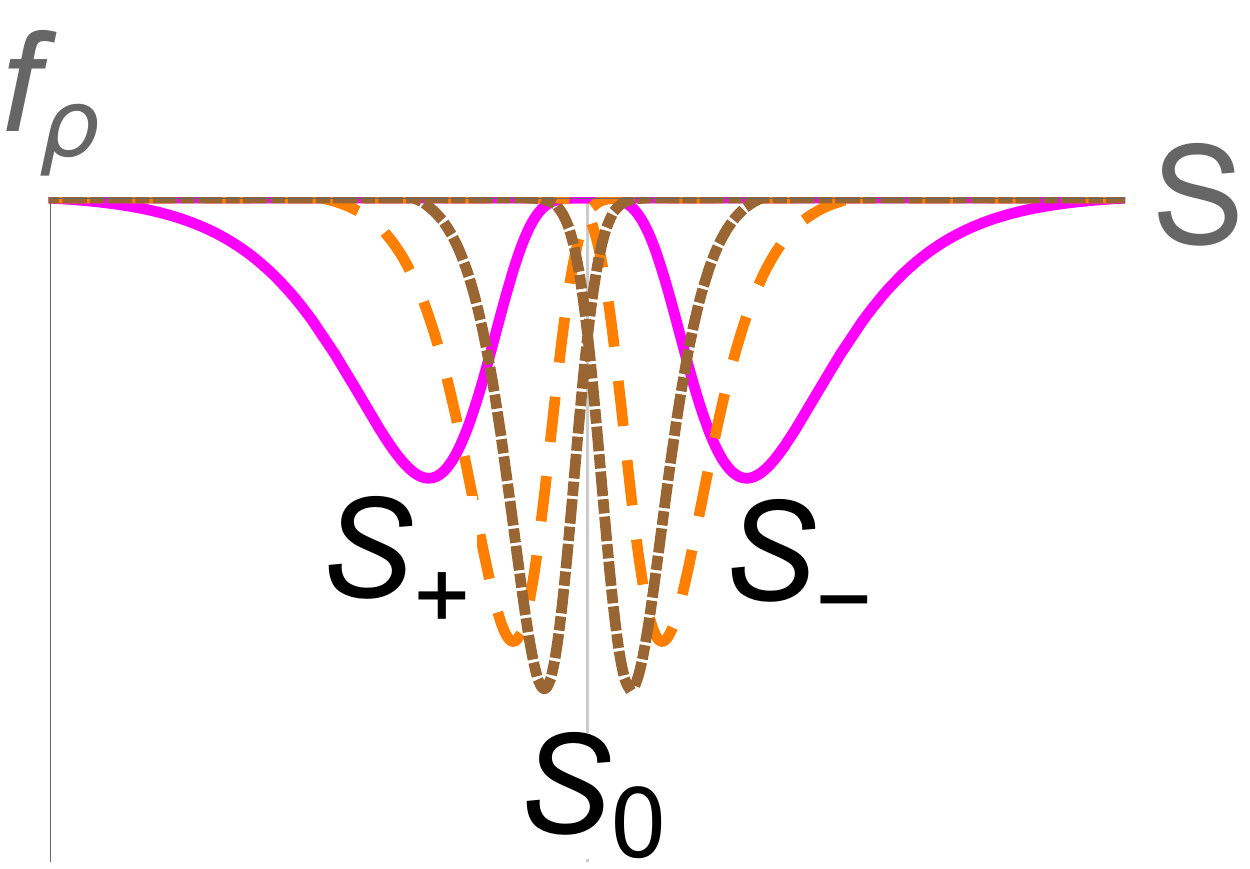}
\label{fig4e}}
\quad
\subfigure[]{\includegraphics[width=.40\columnwidth]{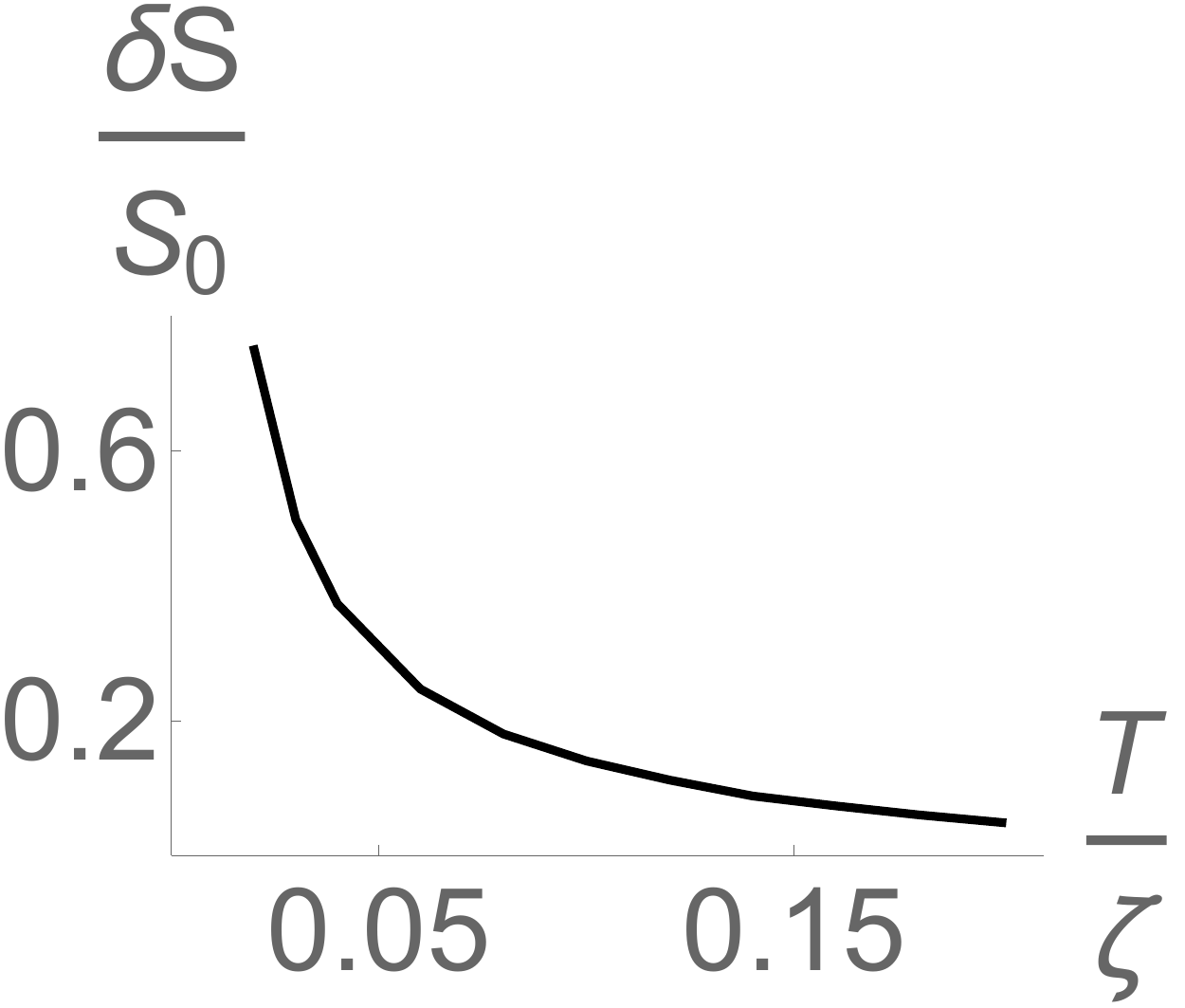}
\label{fig4f}}
\caption{(a) $f_{\rho}$ defined in Eq.~(\ref{dist2}) at different $T$ in metals. The dominant peak is at $\mu$ and does not shift with $T$. Smaller peak height corresponds to higher $T$ (b) $1/m$ vs $E$ for the two bands in Fig.~\ref{fig1}(a) compared to that in metals where it is constant (dashed curve). (c) $f_{\rho}$ for the insulating case in Fig.~\ref{fig1}(a) at different $T$ as a function of $E$. These are obtained by multiplying (b) to (a). Two peaks appear which move farther away with increase in $T$. (d) Area $S$ vs $E$ for the two bands in Fig.~\ref{fig1}(a); lower curve (+): conduction band and the upper curve (-): valence band. (e) Same as in (c) but plotted against $S$ instead of $E$. The peaks move closer to each other with increase in $T$. (f) Difference in the peaks in (e), $\delta S$, as a function of $T$.}
\label{fig4}
\end{figure} 

First, $\Omega_{osc}(T)\approx\Omega^{un}_{osc}(T)$. Since Eq.~(\ref{omegaunzero}) is independent of $E$, on inserting it into Eq.~(\ref{omegatori}) the effect of the integral is simply to introduce an overall prefactor that is $T-$dependent, leaving the frequency unchanged:
\beq
\Omega_{osc}(T)\approx\Omega_{osc}^{un}(T)\propto\mathrm{cos}[S(\tilde{E})l_B^2+\phi].\label{res1}
\eeq 
Thus, $\Omega_{osc}(T)$ oscillates with a single frequency that is same as its frequency at $T=0$ and does not change with $T$. This explains the numerical findings in Fig.~\ref{fig2a}. 
Next, $\rho_{osc}(T)=\rho^c_{osc}(T)$. Inserting Eq.~(\ref{omegaczero}) into Eq.~(\ref{omegatori}), changing the variable of integration from energy $E$ to area $S$ (see Supplementary Materials for details \cite{supp}), and using complex notation for simplicity, I have
\begin{eqnarray}
\rho_{osc}(T)&\propto&Re\left[e^{i\phi}\int f_{\rho}(S,T)e^{iSl_B^2}dS\right],\label{temp2}\\
f_{\rho}(S,T)&=&\frac{1}{m(E)^2}\frac{\partial^2}{\partial\mu^2}\frac{\partial f_0(E,T)}{\partial E}.\label{dist2}
\end{eqnarray}
Above, $f_{\rho}$ is assumed to be expressed in terms of $S$ by inverting the relation $S=S(E)$, and I have used $m(E)=(1/2\pi)dS(E)/dE$.
In spite of the conventional origin, this integral does not behave as in metals. The key difference is that, instead of being a constant as in a metal, $1/m(E)$ in Eq.~(\ref{dist2}) changes rapidly in the interval $[-\zeta,\zeta]$. In Fig.~\ref{fig4a}, I plot $f_{\rho}$ for a standard metal with constant $m$: it is strongly peaked at $\mu$ and the position of the peak does not change with $T$. When the rapidly changing $1/m(E)$, shown in Fig.~\ref{fig4b}, is multiplied to Fig.~\ref{fig4a}, it results in Fig.~\ref{fig4c}. The function is now peaked at two places away from $\mu$ which move farther away from each other with increasing $T$. For both the valence and the conduction bands, the relation between $E$ and $S$ is shown in Fig.~\ref{fig4d}. Using this, $f_{\rho}$ can be replotted in terms of $S$, as shown in Fig.~\ref{fig4e}. Expressed in terms of $S$, the peaks now move closer to each other as $T$ increases. Eq.~(\ref{temp2}) implies averaging an oscillating function over a distribution of frequencies. This results in another oscillating function that has a frequency determined by the position of the peak of the distribution function. Such a reasoning, justified both analytically and numerically in the Supplementary materials \cite{supp}, when applied to Fig.~\ref{fig4d} results in two frequencies given by the areas at which the peaks appear, and they change with $T$, in contrast to metals. Denoting the position of the peaks as $S_{\pm}$, $\rho^c_{osc}(T)\propto \mathrm{cos}[S_+l_B^2+\phi]+\mathrm{cos}[S_-l_B^2+\phi]$. Note, $(S_-+S_+)/2\approx S_0$, where $S_0$ is the area at band intersection prior to hybridization (Fig.~\ref{fig1}). Further, $S_0=S(\tilde{E})$ after hybridization (Fig.~\ref{fig1}). Let $\delta S=(S_--S_+)/2$. Then, I have
\beq
\rho_{osc}(T)\approx\rho^{c}_{osc}(T)\propto \mathrm{cos}[S(\tilde{E})l_B^2+\phi]\mathrm{cos}[\delta S(T)l_B^2].\label{res2} 
\eeq
Thus, $\rho_{osc}(T)$ oscillates with two frequencies that are $T-$dependent. Their manifestation is in the form of a beat with the same basic frequency as $\Omega_{osc}$ [cf. Eq.~(\ref{res1})]  but modulated with a $T-$dependent envelope. This explains the unusual oscillations in Fig.~\ref{fig2b}. Ideally, one should extract the two frequencies from Fig.~\ref{fig2b} by Fourier transform and match them with the predictions of Eq.~(\ref{res2}). Unfortunately, while the beat structure is clearly seen in some curves in Fig.~\ref{fig2b}, in others it is less clear due to the fast decay of oscillations in $1/B$. This, in turn, makes direct extraction of the frequencies difficult. An alternative way is to follow the variation of oscillations at different temperatures but at fixed $1/B$. The factor $\mathrm{cos}[\delta S(T)l_B^2]$ in Eq.~(\ref{res2}) changes sign periodically, predicting a phase flip with change in $T$. Further, from Fig.~\ref{fig4f}, at small $T$, $\delta S(T)$ changes faster, implying more frequent phase flips, compared to that at higher $T$. Both these features are observed in Fig.~\ref{fig2c}, thus validating Eq.~(\ref{res2}). In experiments, a similar approach could be adopted.  

At $T\sim\zeta$, the peaks in Fig.~\ref{fig4e} approach each other and $\delta S\approx 0$ in Eq.~(\ref{res2}). This implies that both $\Omega$ and $\rho$ oscillate with the same single frequency. This is also seen in Figs.~\ref{fig2a} and \ref{fig2b} at the highest temperature.

The results presented here are general and independent of the choice of $\varepsilon_{1,2}(\mb{k})$ in our starting model [Eq.~(\ref{ham})], as long as the system is topologically trivial; non-trivial topology may introduce additional features \cite{zha}. Additionally, the results are valid for both 2D and 3D systems \cite{zha,pal1,pal2}. All the unusual features stem from two key factors unique to a strongly PH asymmetric system: the rapid change of the band curvature within a small interval $[-\zeta,\zeta]$, and the unusual dual origin of oscillations in this interval captured in Eq.~(\ref{omegadec}). Surprisingly, the gap itself plays no role at all! 

It is clear that SdH oscillations are intrinsically much weaker than the dHvA oscillations in a strongly PH asymmetric system. Experimental measurements of SdH oscillations via transport to verify the predictions of this Letter may be challenging since a further reduction in the strength of oscillations will arise due to disorder. A much more direct measurement of the density of states, such as via quantum capacitance or compressibility, could be more suited. Recently, in SmB$_6$, a Kondo insulator that is inherently strongly PH asymmetric, pronounced oscillations in magnetization were observed experimentally, while resistivity showed no oscillations \cite{tan}. Whether these oscillations originate from the bulk or from the surface is currently being intensely debated \cite{tan,li,ert,kno3,den}. The discussion above lends support to the possibility of bulk origin, although this is not conclusive. On the other hand, a more convincing way to distinguish the origin could be to go into the metallic regime such that $\mu$ is near the edge of the band. As discussed before, according to Eq.~(\ref{omegadec}) and the discussion following it, the frequencies of oscillations show unusual behavior in this regime as well. 

A new paradigm for quantum oscillations is emerging where these oscillations can be used to study systems beyond conventional metals. Recent works \cite{kno,zha,pal1} have already shown that oscillations in insulators are qualitatively different from their metallic counterparts: new features appear in the amplitude and the phase. This Letter shows that even within the insulating regime, a strongly PH asymmetric insulator has a \emph{qualitatively} different quantum oscillation footprint compared to a PH symmetric one which shows up in the frequency: unlike in a PH symmetric insulator, in a PH asymmetric insulator dHvA and SdH oscillations show different frequencies, with the frequency of SdH oscillations changing with temperature.

\begin{acknowledgements}
I am grateful to F. Pi{\'e}chon for valuable suggestions at different stages of this work, and to him, J.-N. Fuchs, and G. Montambaux for comments on the manuscript. This work was supported by LabEx PALM Investissement d'Avenir (ANR-10-LABX-0039-PALM).
\end{acknowledgements}


\begin{widetext}

\section{Supplementary Materials}

\section{Comparison of oscillations in Particle-Hole asymmetric and symmetric cases}

The model considered in the main text is that of a strongly particle-hole (PH) asymmetric insulator constructed by hybridizing two overlapping bands $\varepsilon_{1,2}$ with very dissimilar masses $m_{1,2}$, different in sign. The Hamiltonian reads
\beq
H_{\mb{k}}=
\begin{pmatrix}
\ve_1({\mb{k}})-\Delta&\zeta\\
\zeta&\ve_2({\mb{k}})
\end{pmatrix},
\label{ham_supp}
\eeq
where $\Delta$ determines the overlap between the bands before hybridization and $\zeta$ is the hybridizing parameter. In the main text numerical calculations were presented for a lattice model of this Hamiltonian. It was shown that interesting features arise in oscillations in the density of states. It was claimed that this was because of the strong PH asymmetry. Here I present numerically calculated oscillations for the same lattice model for a PH symmetric system and compare them with the asymmetric case. The lack of features in the symmetric case convincingly proves that the features are a result of PH asymmetry.

\begin{figure}[h]
\centering
\subfigure[]{\includegraphics[width=.23\textwidth]{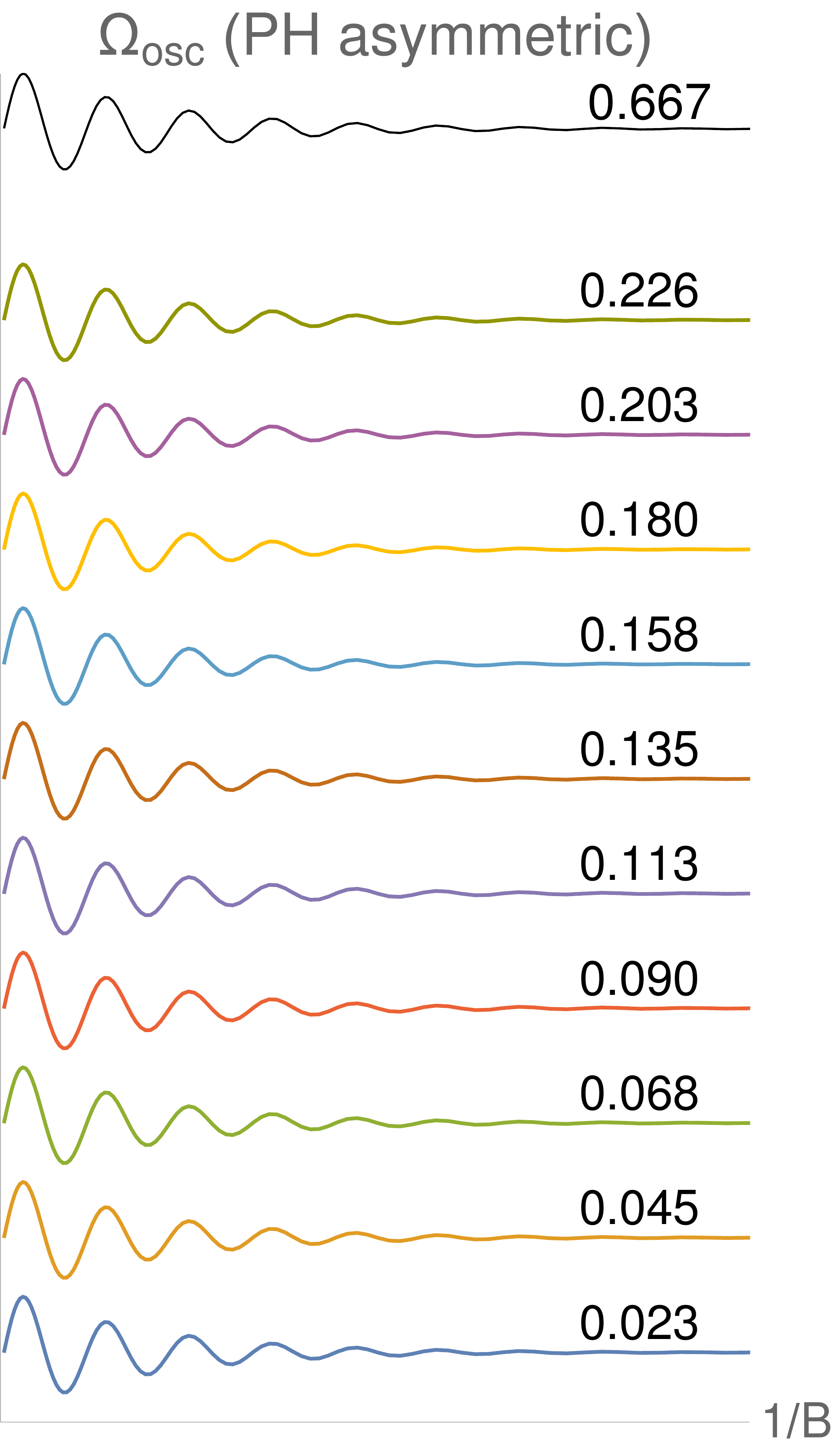}}
\quad
\subfigure[]{\includegraphics[width=.23\textwidth]{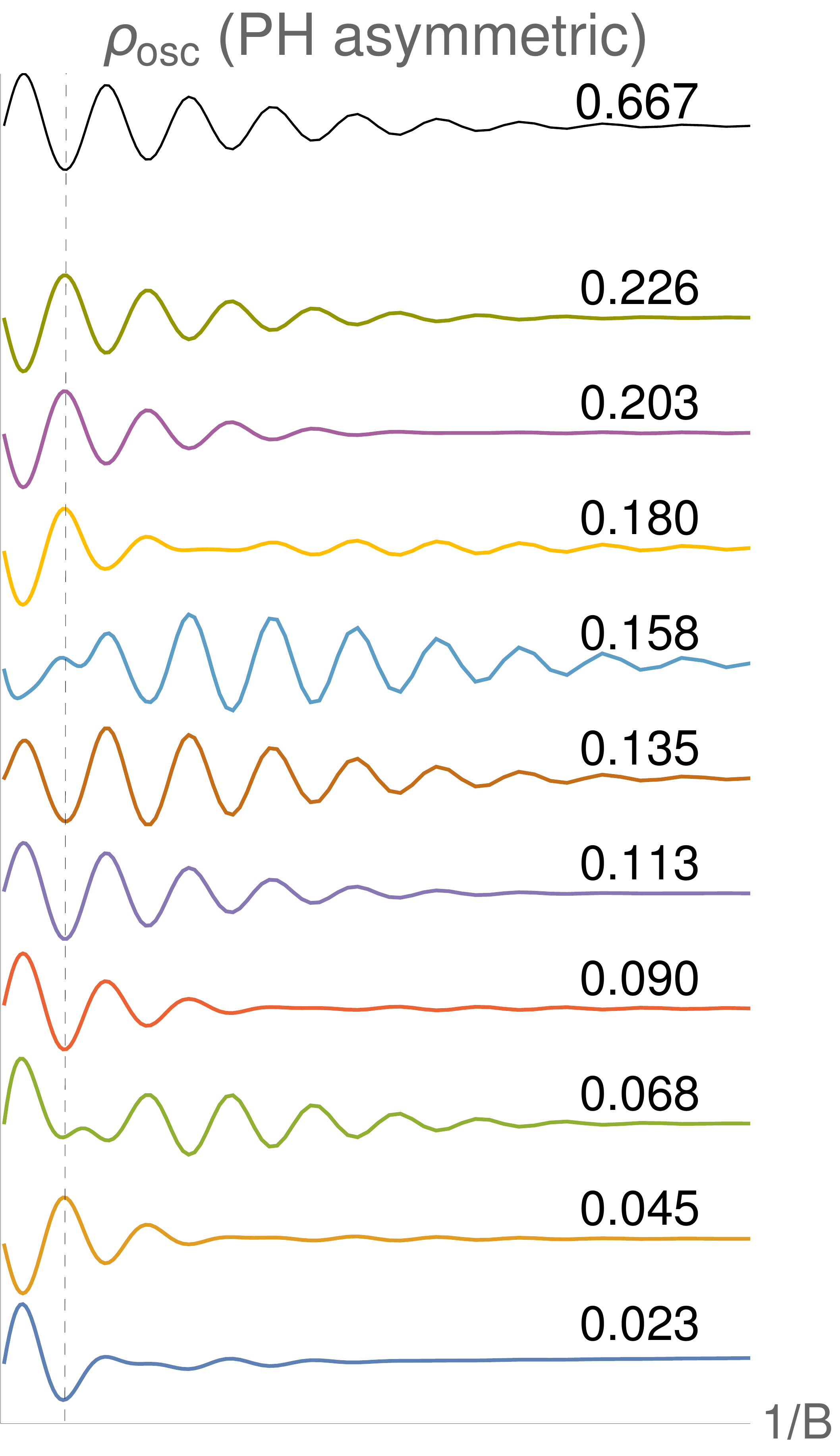}}
\quad
\subfigure[]{\includegraphics[width=.23\textwidth]{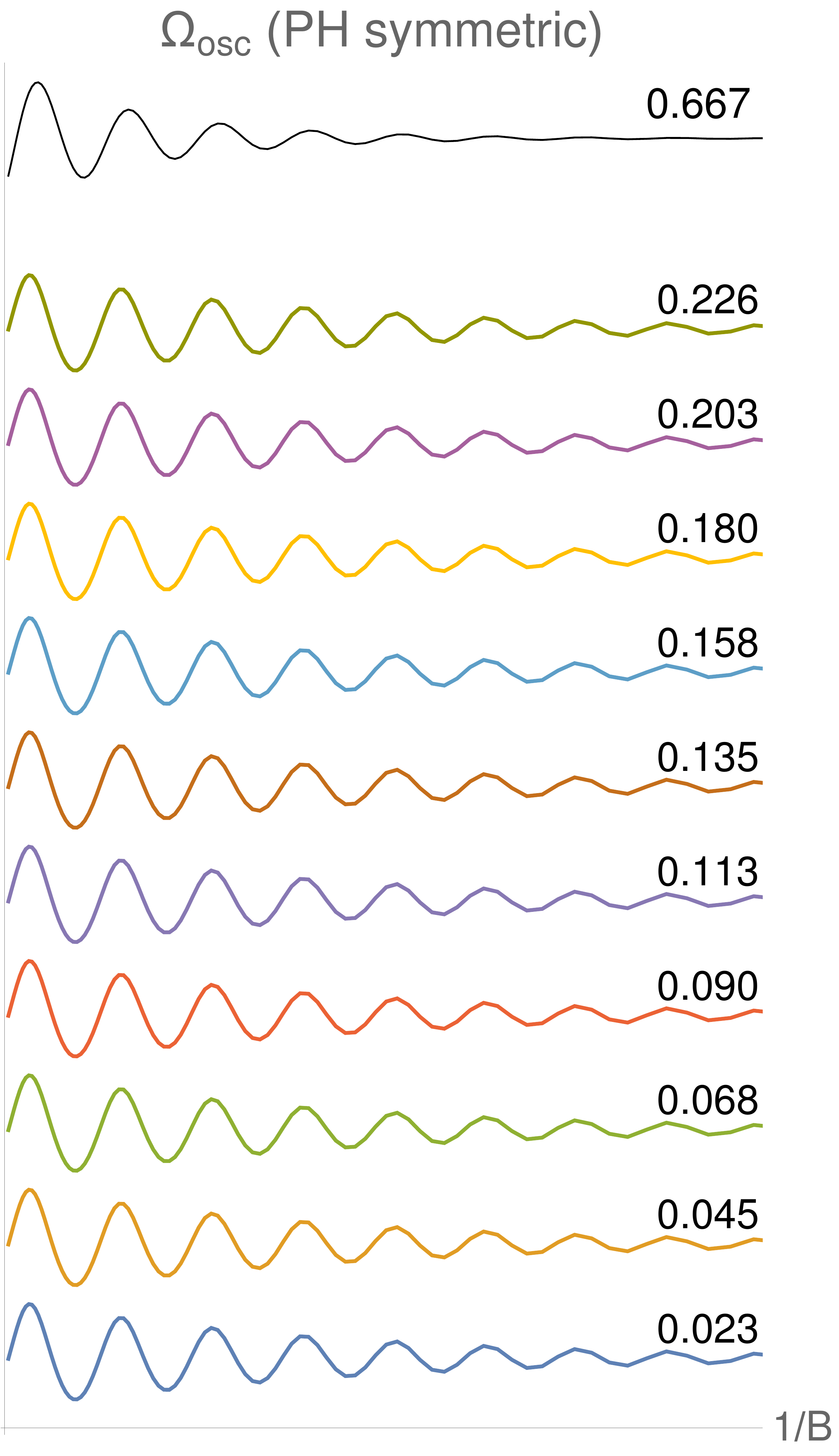}}
\quad
\subfigure[]{\includegraphics[width=.23\textwidth]{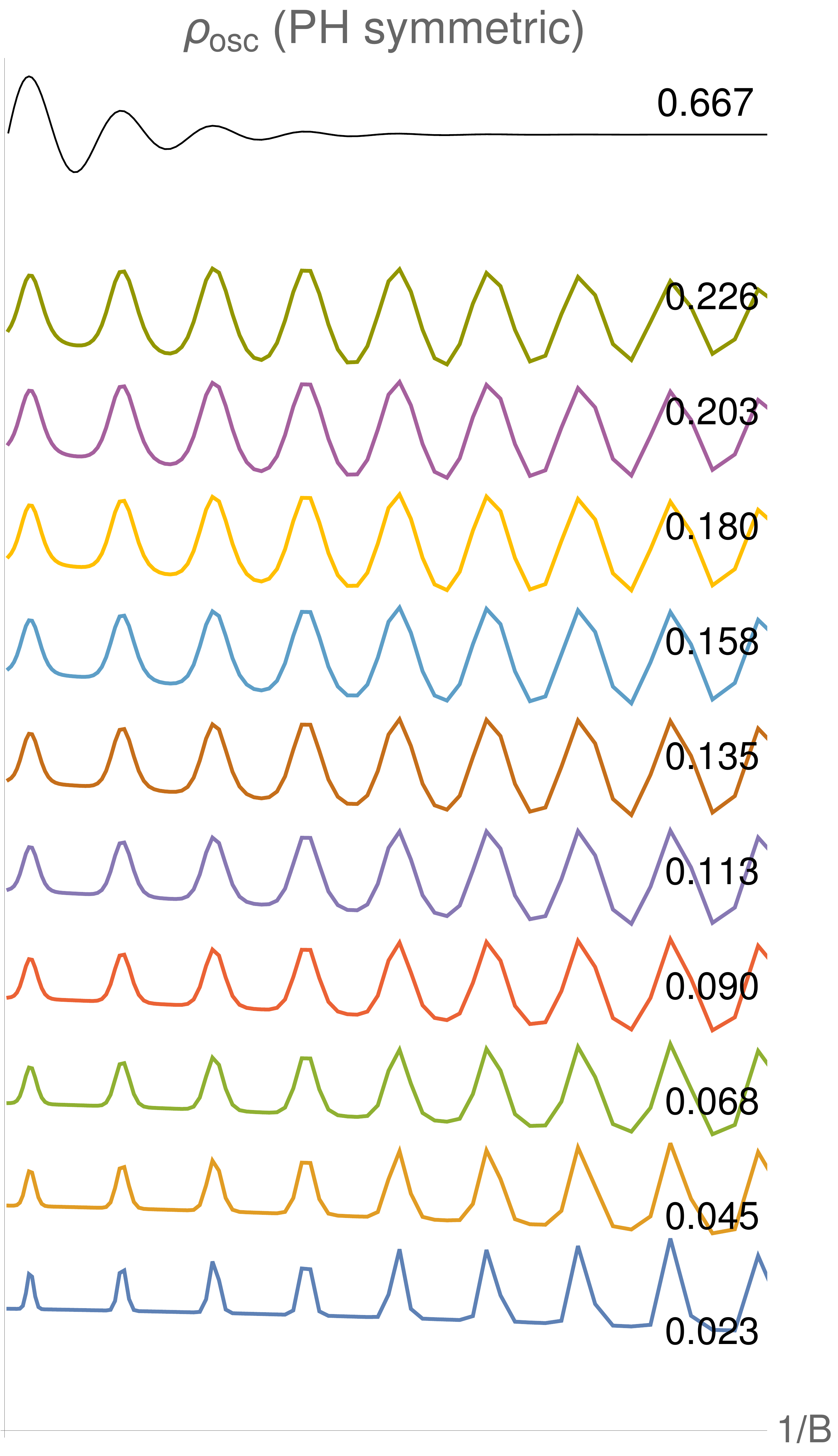}}
\caption{Quantum oscillations numerically calculated for a lattice version of the model in (\ref{ham_supp})---see main text for details. Oscillations in (a) grand potential $\Omega$ and (b) density of states $\rho$ for a strongly PH asymmetric case. Same in (c) and (d) but for a PH symmetric case. The curves are rescaled for clarity such that all have the same amplitude within the field interval studied. Corresponding $T/\zeta$ is provided on the right. It can be seen that the nontrivial features in (b) do not appear in (c) or (d).}
\label{fig0_supp}
\end{figure}

\section{Oscillations near the edge of the band}

\begin{figure}
\includegraphics[angle=0,width=0.50\columnwidth]{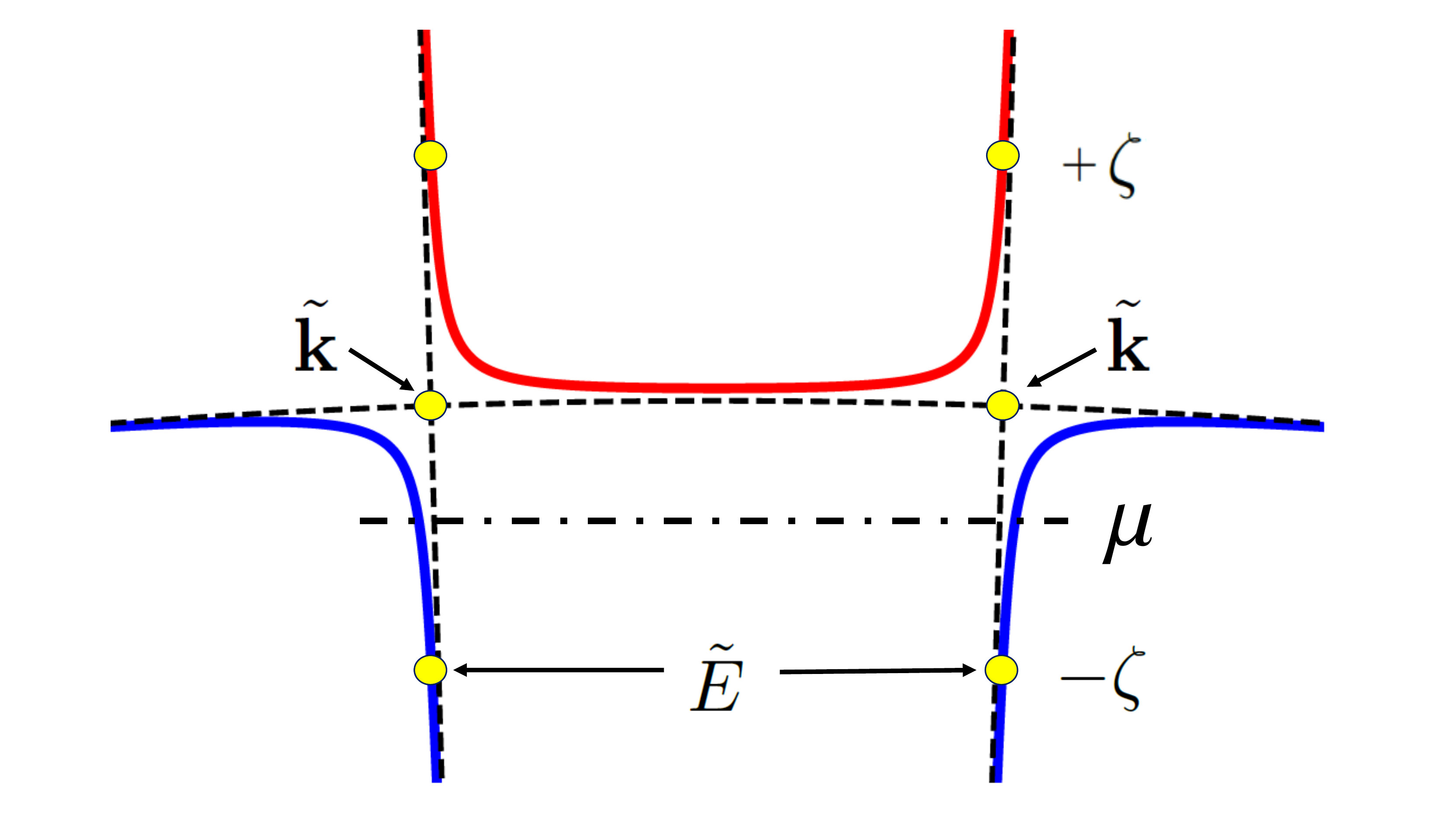}
\caption{Schematic band diagram of a strongly particle-hole asymmetric as studied in the main text. Unhybridized bands  are shown in dashes, intersecting at $\tilde{\mb{k}}$. At $T=0$, when $\mu=0$ is in the gap, oscillations originate from $\tilde{E}=E(\tilde{\mb{k}})=-\zeta$. However, when $\mu$ is in the band region but still within $[-\zeta,\zeta]$, in spite of being in a metallic regime, oscillations do not behave as in a conventional metal. This is addressed in this section.}
\label{fig1_supp}
\end{figure}

When the chemical potential $\mu$ is in the gap, according to Ref.~\cite{pal2_supp}, at zero temperature oscillations in the grand potential $\Omega_{osc}$ arise from $\tilde{E}$ inside the valence band---henceforth referred to as unconventional oscillations, $\Omega^{un}_{osc}$. Here $\tilde{E}=E(\tilde{\mathbf{k}})$, $\tilde{\mathbf{k}}$ being the point of intersection of the two bands before hybridizing. As $\mu$ moves into the band but still lies within $[-\zeta,\zeta]$ (See Fig.~\ref{fig1_supp}), one would expect conventional oscillations arising from $\mu$ as in metals, denoted by $\Omega_{osc}^c$. However, this is not true. In the main text it was argued that in this case, on top of the conventional oscillations arising from $\mu$, the unconventional oscillations arising from $\tilde{E}$, which were there when $\mu$ was in the gap, should also show up . Thus, at $T=0$, inside $[-\zeta,\zeta]$ there are now \emph{two} sources of oscillations: 
\beq
\Omega_{osc}(\mu)=\Omega_{osc}^c(\mu)+\Omega_{osc}^{un}.
\label{omegadec_supp}
\eeq
Here, I present numerical calculations in support of the above equation, and establish Eq.~\ref{omegadec_supp} quantitatively.

\begin{figure}
\includegraphics[angle=0,width=0.99\columnwidth]{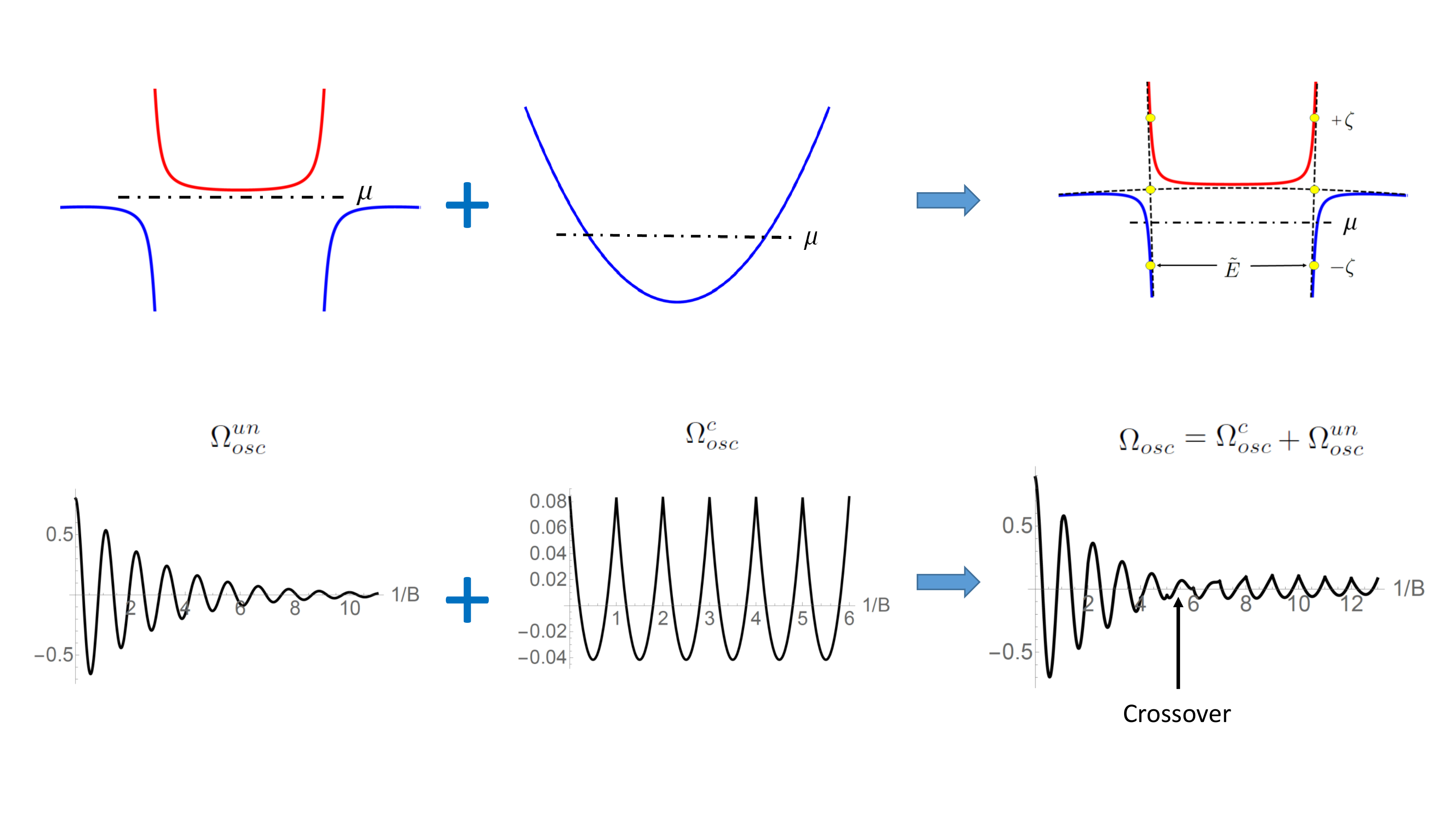}
\caption{Expected oscillation patterns: when $\mu$ is in the gap (left), oscillations ($\Omega_{osc}^{un}$) are smooth and decay as $1/B$ increases. In a normal metal  (middle), oscillations ($\Omega_{osc}^c$) arising from $\mu$ are sharp and they do not decay with $1/B$. In Eq.~\ref{omegadec_supp} it has been claimed that when $\mu$ is near the edge of the band, i.e., inside the band but still within $[-\zeta,\zeta]$, even though it is in the metallic regime, oscillations are not simply that of a metal---it has both the conventional contribution from $\mu$ and the contribution that was there when $\mu$ was in the gap. The resulting oscillation pattern is shown in the right. At smaller values of $1/B$, $\Omega_{osc}^{un}$ wins but at higher values of $1/B$ it decays and $\Omega_{osc}^c$ wins. Since the amplitude of $\Omega_{osc}^c\propto 1/m(\mu)$ (see Eq.~\ref{omegaczero_supp}), it is weaker as one goes closer to the edge of the band. Accordingly, the crossover region between the two contributions will also move to the right as one moves towards the edge. Such expected features are to be compared with Fig.~\ref{fig3_supp} which is obtained by exact numerical calculations.}
\label{fig2_supp}
\end{figure}

Before presenting the numerical calculations, I present in Fig.~\ref{fig2_supp} the theoretically expected oscillation pattern following Eq.~(\ref{omegadec_supp}). As discussed in Refs.~\cite{pal2_supp,kno_supp}, the contribution $\Omega^{un}_{osc}$ arising from $\tilde{E}$ is a smooth oscillating curve that decays as $1/B$ increases. The oscillation frequency $F^{un}\propto S(\tilde{E})$ where $S(\tilde{E})$ is the area of the orbit in $k$-space at energy $\tilde{E}$. One can then describe it as
\beq
\Omega_{osc}^{un}(\mu)\propto g\left(\frac{\zeta}{\omega_c}\right)\mathrm{cos}[S(\tilde{E})l_B^2+\phi], \label{omegaunzero_supp}
\eeq
where $g(\zeta/\omega_c)$ is a function describing the decay of the amplitude whose exact form is not required ($\omega_c=eB/m^0$ with $m^0$ the band mass of the lighter unhybridized band), and $l_B^2=1/eB$ is the magnetic length squared. Note that this contribution is completely independent of $\mu$ as long as it is within $[-\zeta,\zeta]$. Outside of this region, it is zero. In contrast, the contribution $\Omega^c_{osc}$ arising from $\mu$ is the regular metallic contribution. This does not decay with $1/B$ and it has a sharp waveform as compared to $\Omega^{un}_{osc}$; see Shoenberg \cite{sho_supp} for a discussion. It can be described as
\beq
\Omega_{osc}^c(\mu)\propto\frac{1}{|m(\mu)|} \mathrm{cos}[S(\mu)l_B^2+\phi],\label{omegaczero_supp}
\eeq
where $m(\mu)$ is the band mass at $\mu$. This contribution varies as $\mu$ varies, both in frequency $F^c\propto S(\mu)$ and in amplitude $\propto 1/|m(\mu)|$. Importantly, the two contributions are not of equal strength: for $\zeta/\omega_c\ll 1$ one can expand $g(\zeta/\omega_c)$ to get $|\Omega^{un}_{osc}|\sim|\Omega_{osc}^0|[1-\mathcal{O}(\zeta/\omega_c)]$. On the other hand, $|\Omega^c_{osc}|\sim|\Omega_{osc}^0||m^0/m(\mu)|$, where the superscript $0$ denotes quantities prior to hybridization ($m^0$ is the lighter unhybridized mass). At smaller values of $1/B$, $\Omega^{un}_{osc}$ wins until $g\approx m^0/m$, where a crossover happens and beyond this, at higher values of $1/B$, $\Omega^{un}_{osc}$ dies leaving only $\Omega^c_{osc}$. This leads to the unique oscillation pattern shown in Fig.~\ref{fig2_supp}: the two contributions with different frequencies and waveform dominate at different regions of $1/B$; in numerics they can, therefore, be easily identified.

\begin{figure}
\includegraphics[angle=0,width=\columnwidth]{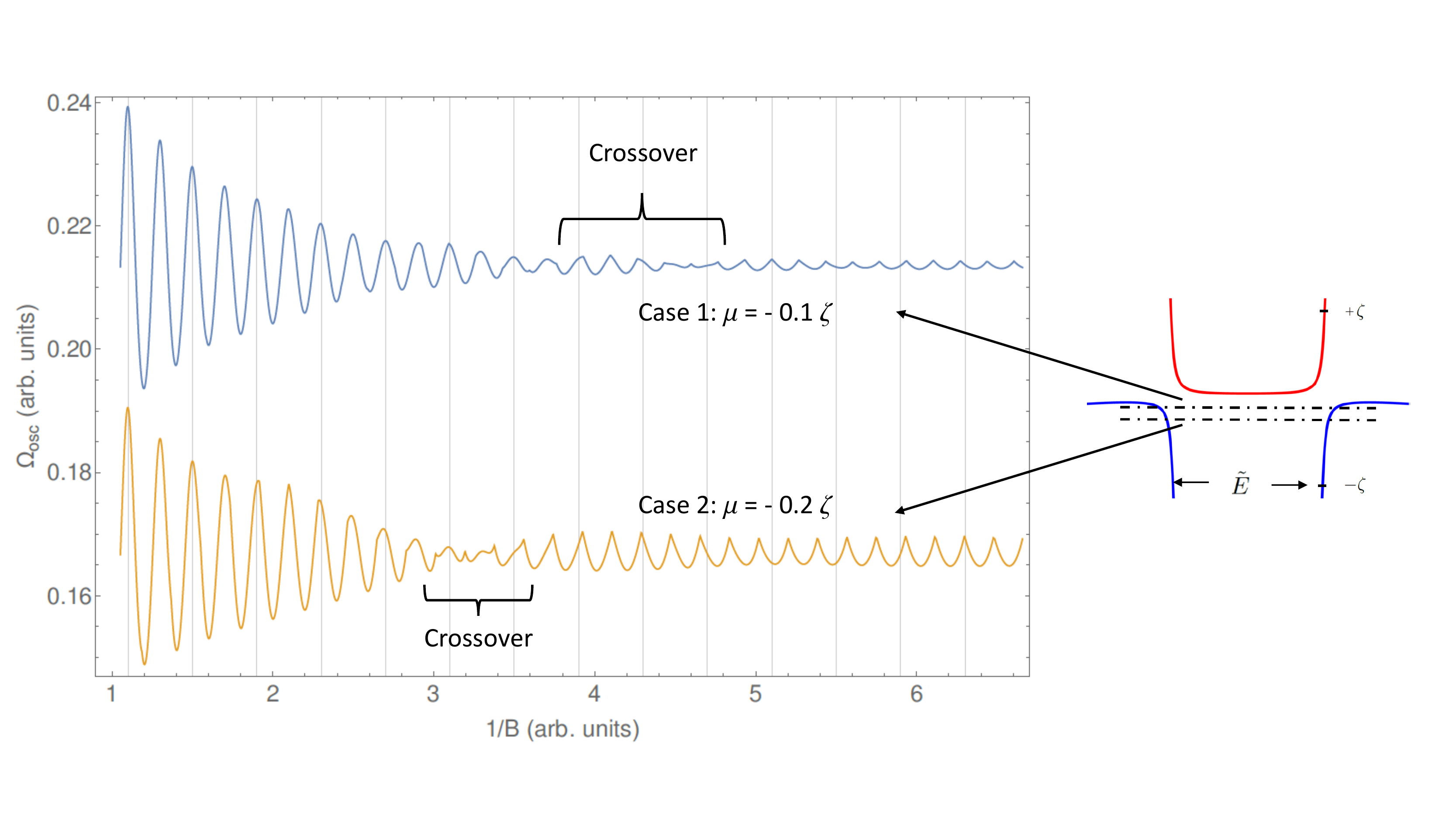}
\caption{Numerically calculated oscillations in the grand potential, $\Omega_{osc}$, for the model in Eq.~\ref{ham_supp}. I have used $\varepsilon_{1,2}(\mb{k})=k^2/m_{1,2}$ with $m_1/m_2=-0.0001$ and $\zeta/\Delta=0.02$. Oscillations are presented for two values of $\mu$: $\mu=-0.1 \zeta$ (case 1) and $\mu=-0.2 \zeta$ (case 2). The oscillation pattern is exactly similar to the theoretically expected pattern shown in Fig.~\ref{fig2_supp}. At lower values of $1/B$ oscillations are due to $\Omega_{osc}^{un}$ which do not change as $\mu$ is moved, as predicted. On the other hand, at higher values of $1/B$, oscillations are due to $\Omega_{osc}^c$ where both the amplitude and the frequency change as $\mu$ is moved. It can be seen that the amplitude decreases and the crossover region moves to the right as $\mu$ moves towards the edge of the band, agreeing with the theoretical prediction. The frequencies of the two contributions can be extracted and compared---see Table \ref{tab1}.}
\label{fig3_supp}
\end{figure} 

\begin{table*}[h]
\centering
\caption{\label{tab1} Frequency $F^c$, extracted from the right side of the crossover region in Fig.~\ref{fig3_supp}, corresponds to $\Omega_{osc}^c$ while $F^{un}$, extracted from the left side of the crossover region, corresponds to $\Omega_{osc}^{un}$. Area of orbit in $k$-space for the model used in Fig.~\ref{fig3_supp} have been calculated. According the theory presented here, $F^c\propto S(\mu)$ and $F^{un}\propto S(\tilde{E})$. Comparing the ratio of $F^c/F^{un}$ with $S(\mu)/S(\tilde{E})$, an excellent agreement is found.}
\begin{tabular}{ccc}
\hline\noalign{\smallskip}
\hline
\noalign{\smallskip}
$\mu$ & $F^c/F^{un}$ & $S(\mu)/S(\tilde{E})$\\  \noalign{\smallskip} \hline
\hline\noalign{\smallskip}
$-0.1 \zeta$  (case 1) & 1.195  & 1.200 \\\noalign{\smallskip}\hline
\noalign{\smallskip}
$-0.2 \zeta$  (case 2) & 1.099 & 1.096 \\\noalign{\smallskip}\noalign{\smallskip}\hline
\end{tabular}
\end{table*}

With the above in mind, I now present actual numerical calculations in Fig.~\ref{fig3_supp} for the model described in Eq.~\ref{ham_supp}. For the calculations I have chosen $\varepsilon_{1,2}(\mb{k})=k^2/m_{1,2}$ with $m_1/m_2=-0.0001$. I present oscillations in the grand potential $\Omega_{osc}$ for two values of $\mu$: (1) $\mu=-0.1 \zeta$ and (2) $\mu=-0.2 \zeta$. It can be immediately seen that both the curves match very well with the theoretically expected curve in Fig.~\ref{fig2_supp}. Indeed, both curves begin as smooth decaying oscillations which cross over to a pattern that has a sharp non-decaying waveform. The crossover region in case 2 moves to the left compared to case 1. This is expected since this region happens when $g(\zeta/\omega_c)\approx m_0/m(\mu)$, and $m(\mu)$ in case 1 is higher than in case 2. The frequency at lower values of $1/B$ does not change as $\mu$ moves confirming that this arises from $\tilde{E}$. On the other hand, the frequency at higher values of $1/B$ does change as $\mu$ moves confirming that this is the regular contribution. More quantitatively, one expects the ratio of the frequencies of the conventional and unconventional parts to be equal to the ratio of the areas from where these originate, i.e., $F^c/F^{un}=S(\mu)/S(\tilde{E})$. In Table \ref{tab1} I compare these for the two cases. The frequencies are extracted from the numerical curves in Fig.~\ref{fig3_supp} and the areas are calculated theoretically for the model used. The excellent quantitative agreement validates the claim made in Eq.~(\ref{omegadec_supp}), a central point used in the main text.

\section{Temperature dependence of frequency of oscillations in the density of states}

In the main text, based on qualitative explanations and intuitive arguments, I explained why the frequency of oscillations in the density of states (DOS) changes with temperature. Here, I justify those arguments by providing analytical and numerical calculations.

The grand potential at nonzero temperature is given by:
\beq
\Omega_{osc}(\mu,T)=\int_{-\infty}^{\infty}\frac{-\partial f_0(E-\mu,T)}{\partial E}\Omega_{osc}(E,0)dE,
\label{omegatori_supp}
\eeq
where $f_0$ is the Fermi-Dirac function. In the previous section it was shown that $\Omega$ consists of two parts in the region $[-\zeta,\zeta]$: a conventional part and an unconventional part. As explained in the main text, the latter does not contribute to oscillations at $T<\zeta$, only the former does. Inserting the expression from Eq.~(\ref{omegaczero_supp}) in Eq.~(\ref{omegatori_supp}), using the definition $\rho=-\partial^2\Omega/\partial\mu^2$, and employing the complex notation for simplicity, I have
\begin{eqnarray}
\rho_{osc}(\mu,T)&\propto& Re\left[e^{i\phi}\int_{-\infty}^{\infty}\bar{f}_{\rho}(E-\mu,T)e^{iS(E)l_B^2}dE\right]\nonumber\\
&=& Re\left[e^{i\phi}\left\{\underbrace{\int_{-\infty}^{0}\bar{f}_{\rho}(E-\mu,T)e^{iS(E)l_B^2}dE}_{\mathrm{valence\ band}}+\underbrace{\int_{0}^{\infty}\bar{f}_{\rho}(E-\mu,T)e^{iS(E)l_B^2}dE}_{\mathrm{conduction\ band}}\right\}\right].
\label{temp2_supp}
\end{eqnarray}
with
\begin{equation}
\bar{f}_{\rho}(E-\mu,T)=\frac{1}{|m(E)|}\frac{\partial^2}{\partial\mu^2}\frac{\partial f_0(E-\mu,T)}{\partial E}.\label{dist2_supp}
\end{equation}
This is formally same as in a conventional metal.  In spite of this, the effect of temperature for the insulating case described by Eq.~\ref{ham_supp} is different from that in a metal. In a conventional metal, $f_{\rho}$ is peaked at $E=\mu$ and this position does not change with temperature. One can then expand $S(E)$ near $\mu$ as $S(E)\approx  S(\mu)+S'(\mu)(E-\mu)$ and inserting it in Eq.~\ref{temp2_supp}, the integral can be evaluated to get the standard Lifshitz-Kosevich result. Note that, in Eq.~\ref{dist2_supp}, $m(E)=m$ does not vary on the scale of $T$ and is a constant. In the case of the insulator considered here, $S(E)$ [and consequently $m(E)=(1/2\pi) dS/dE$] changes rapidly, and, therefore, it can not be expanded as in metals. Instead, we make a change of variables to rewrite Eq.~\ref{temp2_supp} as \cite{note1}
\beq
\rho_{osc}(\mu,T)\propto Re\left[e^{i\phi}\left\{\underbrace{\int_{0}^{\infty}f_{\rho-}(S)e^{iSl_B^2}dS}_{\mathrm{valence\ band}}+\underbrace{\int_{0}^{\infty}f_{\rho+}(S)e^{iSl_B^2}dS}_{\mathrm{conduction\ band}}\right\}\right],
\label{temp2n_supp}
\eeq
with
\begin{equation}
f_{\rho\pm}=\frac{1}{m(E)^2}\frac{\partial^2}{\partial\mu^2}\frac{\partial f_0}{\partial E}.\label{dist2n_supp}
\end{equation}
for the respective bands. Above we have used the fact that $dS(E)/dE=2\pi m(E)$ and all functions of $E$ are assumed to be written in terms of $S$ by inverting the relation $S=S(E)$. The function $f_{\rho\pm}$ is still strongly peaked as in the metallic case, but now, because $m(E)$ changes rapidly within $[-\zeta,\zeta]$, it shows a different qualitative behavior: it has peaks \emph{away} from $\mu$ and their positions shift with temperature as shown in Fig.~\ref{fig4_supp}. Denoting the peak positions as $S_{\pm}$ in the valence and conduction band, respectively, one can rewrite
\begin{eqnarray}
\rho_{osc}(\mu,T)&\propto&Re\left[e^{i\phi}\left\{e^{iS_+(T)l_B^2}\int_{0}^{\infty}f_{\rho-}(S)e^{i(S-S_+)l_B^2}dS+e^{iS_-(T)l_B^2}\int_{0}^{\infty}f_{\rho+}(S)e^{i(S-S_-)l_B^2}dS\right\}\right]\nonumber\\
&=&Re\left[e^{i\phi}\left\{e^{iS_+(T)l_B^2}\int_{-\infty}^{\infty}f_{\rho-}(x)e^{ix}dx+e^{iS_-(T)l_B^2}\int_{-\infty}^{\infty}f_{\rho+}(x)e^{ix}dx\right\}\right]
\label{temp2nn_supp}
\end{eqnarray}
The two integrals are equal except for a phase which are equal but opposite from symmetry. Denoting them as $\mathcal{I}e^{i\mathcal{\pm\theta}}$, I have
\begin{eqnarray}
\rho_{osc}(\mu,T)&\propto&Re\left[e^{i\phi}\mathcal{I}\left\{e^{iS_+(T)l_B^2+\theta}+e^{iS_-(T)l_B^2-\theta}\right\}\right]\\
&\propto&\mathrm{cos}[S_0l_B^2+\phi]\mathrm{cos}[\delta S(T)l_B^2+\theta(T)],
\end{eqnarray}
where $S_0=(S_++S_-)/2$ and $\delta S=(S_--S_+)/2$. Looking at Fig.~\ref{fig1_supp}, $S_0$ is approximately the area at the intersection of the bands prior to hybridization and is equal to $S(\tilde{E})$ after hybridization. Additionally, $\delta S(T)$ changes much rapidly than $\theta (T)$; therefore, the latter can be neglected. These considerations lead to 
\beq
\rho_{osc}(\mu,T)\propto\mathrm{cos}[S(\tilde{E})l_B^2+\phi]\mathrm{cos}[\delta S(T)l_B^2].
\label{mainT_supp}
\eeq
This result is quoted in the main text. The upshot is that oscillations in the DOS comprises two frequencies which change with temperature. And, these frequencies can simply be read off by figuring out the peak position of $f_{\rho}(S)$ as temperature changes. A further quantitative proof of the latter statement is presented in Table~\ref{tab2}. The integral in Eq.~\ref{temp2n_supp} can be calculated numerically and the frequency can be extracted from the oscillations at different temperatures. I do this for the valence band side and compare them with the frequencies expected from the peak positions of $f_{\rho}(S)$. It can be seen that they match quite well, thus lending support to Eq.~\ref{mainT_supp}.

\begin{figure*}
\centering
\subfigure[]{\includegraphics[width=.23\textwidth]{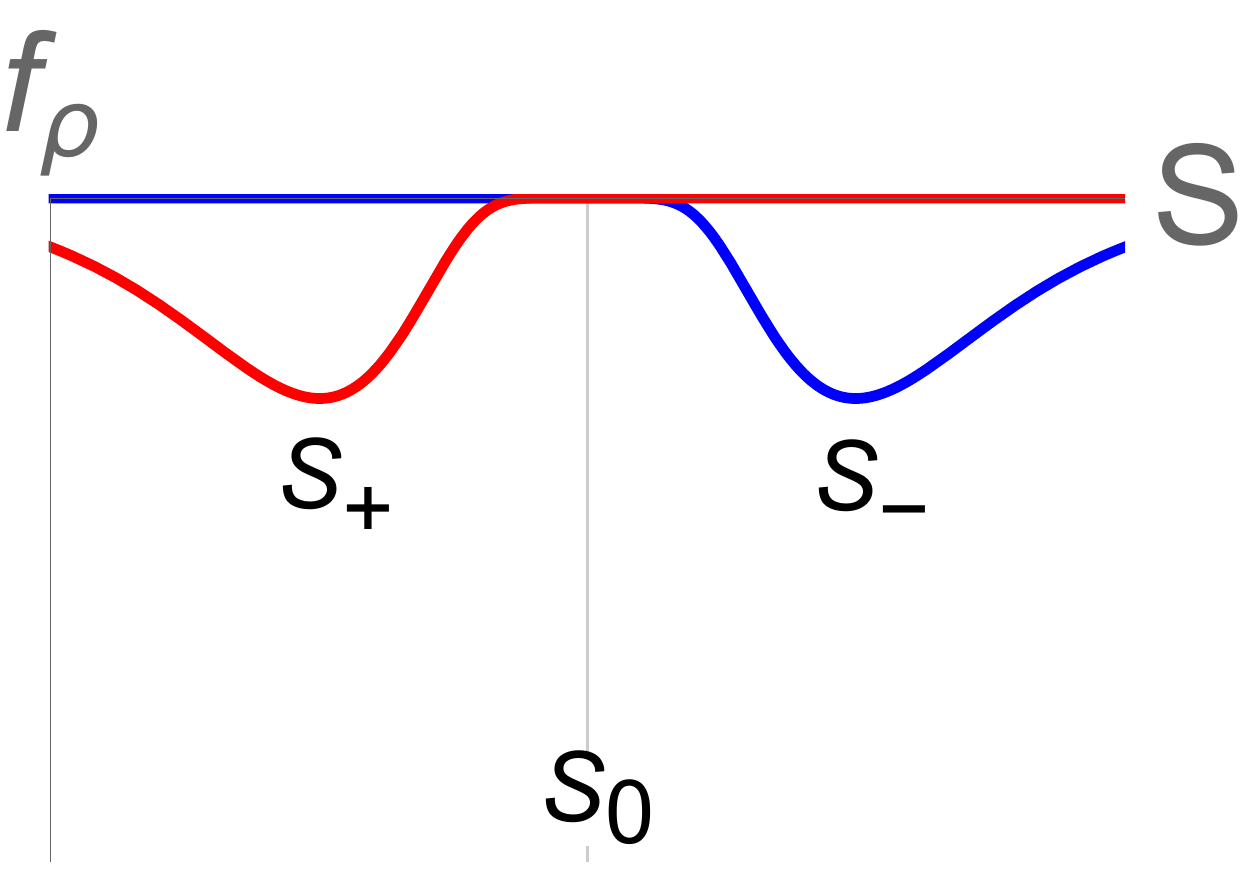}}
\quad
\subfigure[]{\includegraphics[width=.23\textwidth]{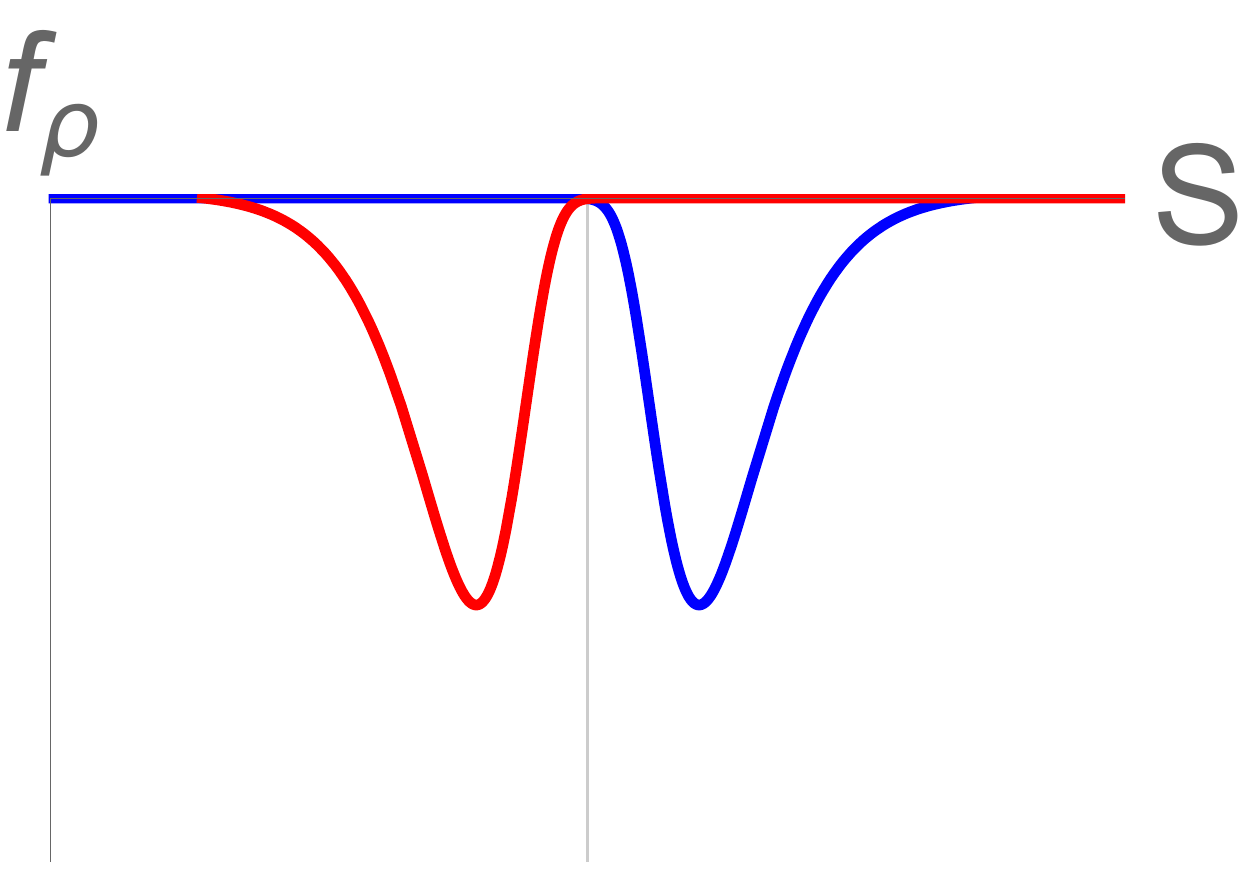}}
\quad
\subfigure[]{\includegraphics[width=.23\textwidth]{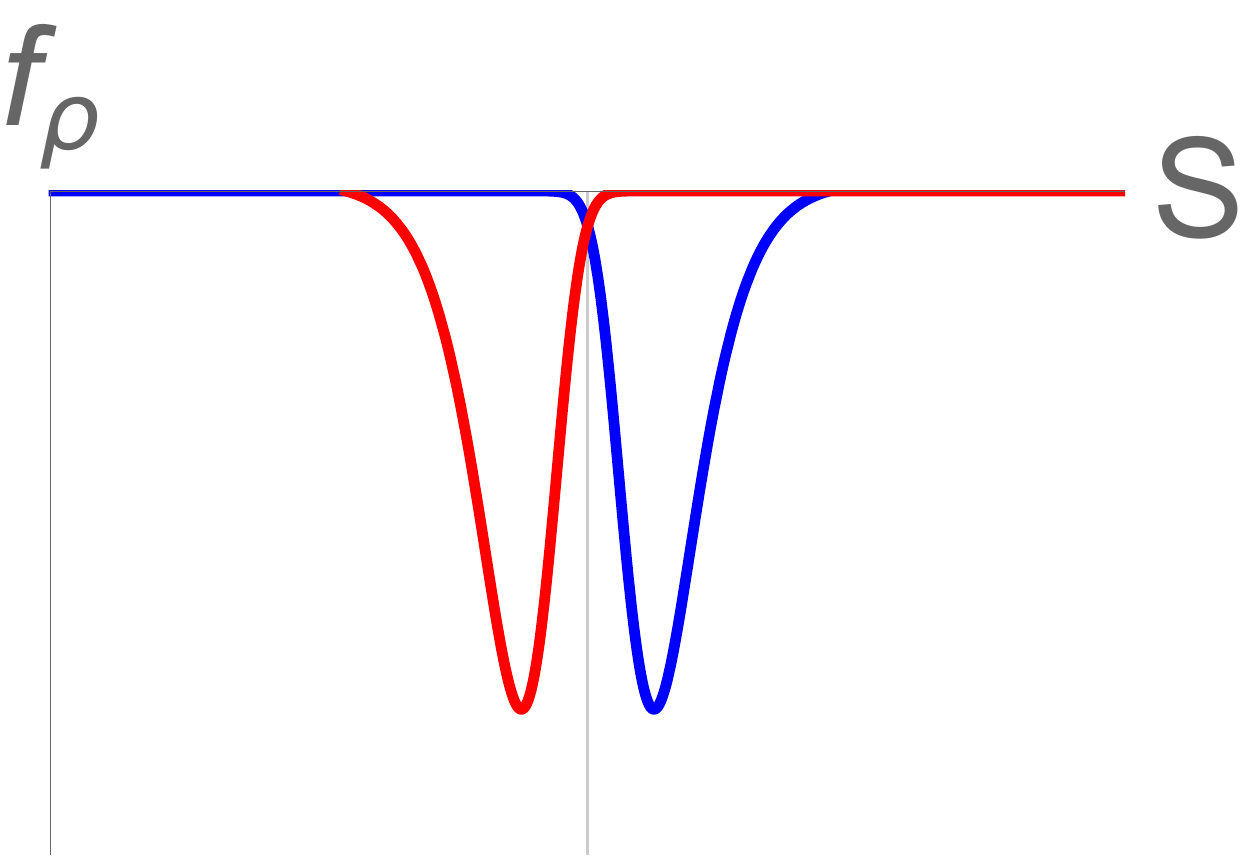}}
\quad
\subfigure[]{\includegraphics[width=.23\textwidth]{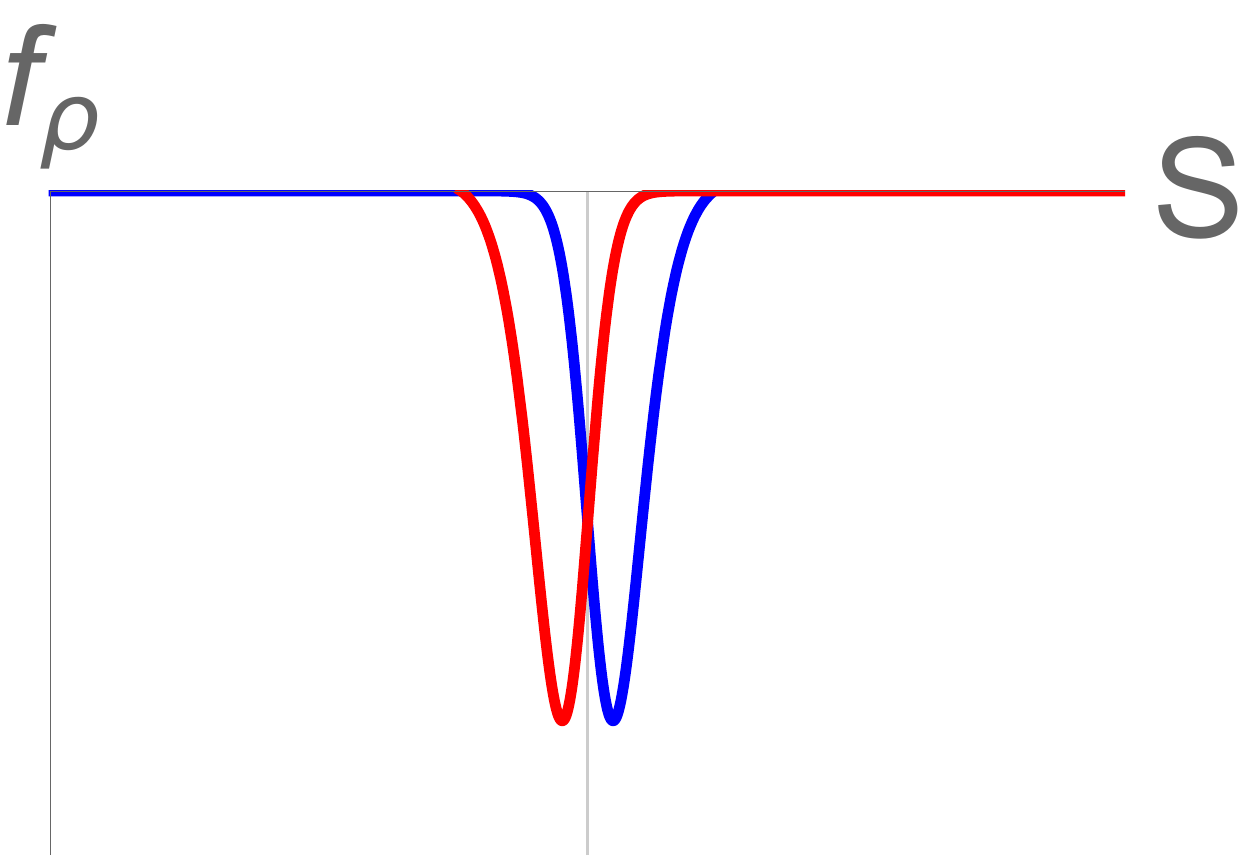}}
\caption{The function $f_{\rho\pm}$ defined in Eq.~\ref{dist2n_supp} as a function of the area $S$. The red curve is for the conduction band ($f_{\rho+}$) and the blue curve is for the valence band ($f_{\rho+}$). Temperature increases as one goes to the right: $T/\zeta=0.03, 0.07,0.11,0.20$, respectively.}
\label{fig4_supp}
\end{figure*}

\begin{table*}
\centering
\caption{\label{tab2} Frequency in column 2 is found by calculating the integral in Eq.~\ref{temp2n_supp} numerically and extracting the frequency (normalized by the frequency of oscillations in the unhybridized case). In column 3, I present the position of the peak as found in Fig.~\ref{fig4_supp}. I have only considered only the valence band side here. According to the theory presented here, $F_-(T)\propto S_-(T)$ and one expects $F_-(T)/F_0=S_-(T)/S_0$. This is verified here.}
\begin{tabular}{ccc}
\hline\noalign{\smallskip}
\hline
\noalign{\smallskip}
$T/\zeta$ & $F_-/F^0$ & $S_-/S_0$\\  \noalign{\smallskip} \hline
\hline\noalign{\smallskip}
$0.05$ & 1.146  & 1.130 \\\noalign{\smallskip}\hline
\noalign{\smallskip}
$0.1$ & 1.068 & 1.082 \\\noalign{\smallskip}\noalign{\smallskip}\hline
\noalign{\smallskip}
$0.5$ & 0.973 & 0.980 \\\noalign{\smallskip}\noalign{\smallskip}\hline
\end{tabular}
\end{table*}


\end{widetext}


\begin{thebibliography}{99}


\bibitem{sho} D. Shoenberg, \emph{Magnetic Oscillations in Metals}, Cambridge Univ. Press (1984).


\bibitem{tan} B. S. Tan, Y. -T. Hsu, B. Zeng, M. Ciomaga Hatnean, N. Harrison, Z. Zhu, M. Hartstein, M. Kiourlappou, A. Srivastava, M. D. Johannes, T. P. Murphy, J. -H. Park, L. Balicas, G. G. Lonzarich, G. Balakrishnan, S. E. Sebastian, Science \tb{349}, 287 (2015).



\bibitem{kno} J.  Knolle and Nigel R. Cooper, Phys. Rev. Lett. \tb{115}, 146401 (2015).
 
\bibitem{zha} L. Zhang, X. Song, and F. Wang, Phys. Rev. Lett. \tb{116}, 046404 (2016).

\bibitem{pal1} H. K. Pal, F. Pi{\'e}chon, J-N. Fuchs, M. Goerbig, and G. Montambaux, Phys. Rev. B \tb{94}, 125140 (2016). 

\bibitem{pal2} H. K. Pal,  Phys. Rev. B \tb{95}, 085111 (2017).

\bibitem{kno2} J. Knolle and N. R. Cooper, Phys. Rev. Lett. \tb{118}, 176801 (2017).

\bibitem{kum} P. Ram and B. Kumar, arXiv:1702.02825 (2017).

\bibitem{fri} S. Grubinskas and L. Fritz, arXiv:1704.06403 (2017).

\bibitem{bal}  J. Liu and L. Balents, Phys. Rev. B \tb{95}, 075426 (2017).

\bibitem{rus} Z. Z. Alisultanov, JETP Lett. \tb{104}, 188 (2016).


\bibitem{supp} [URL will be inserted by publisher]

\bibitem{li} G. Li, Z. Xiang, F. Yu, T. Asaba, B. Lawson, P. Cai, C. Tinsman, A. Berkley, S. Wol-
gast,  Y. S. Eo,  D. -J. Kim,  C ̧ . Kurdak,  J. W. Allen,  K. Sun,  X. H. Chen,  Y. Y. Wang,
Z. Fisk, L. Li, Science \tb{346}, 1208 (2014).

\bibitem{ert} O. Erten, P. Ghaemi, and P. Coleman, Phys. Rev. Lett. \tb{116}, 046403 (2016).

\bibitem{kno3} J. Knolle and N. R. Cooper, Phys. Rev. Lett. \tb{118}, 096604 (2017).
 
\bibitem{den} J. D. Denlinger, Sooyoung Jang, G. Li, L. Chen, B. J. Lawson, T. Asaba, C. Tinsman, F. Yu, Kai Sun, J. W. Allen, C. Kurdak, Dae-Jong Kim, Z. Fisk, and Lu Li, arXiv:1601.07408v1 (2016).


\end{thebibliography}

\begin{thebibliography}{99}


\bibitem{pal2_supp} H. K. Pal,  Phys. Rev. B \tb{95}, 085111 (2017).

\bibitem{kno_supp} J.  Knolle and Nigel R. Cooper, Phys. Rev. Lett. \tb{115}, 146401 (2015).
 

\bibitem{sho_supp} D. Shoenberg, \emph{Magnetic Oscillations in Metals}, Cambridge Univ. Press (1984).

\bibitem{note1} I thank F. Pi\'{e}chon for suggesting me this step.


\end{thebibliography}
\end{document}